\documentclass[10pt,aps,prl,twocolumn,superscriptaddress]{revtex4-1}

\usepackage{amsmath}
\usepackage[utf8]{inputenc}
\usepackage{braket}
\usepackage{mathtools}

\usepackage{hyperref}
\usepackage{lipsum}
\usepackage{xcolor}
\usepackage{comment}
\usepackage{bm,amssymb,graphicx,braket,dsfont,mathtools,bbold,MnSymbol}

\usepackage[caption=false]{subfig}
\usepackage{pgfplots}
\usepackage{cancel}
\usepackage{tabularx,booktabs}
\usepackage{multirow}
\usepackage{fancyhdr}
\usepackage{amsthm}
\usepackage{inputenc}
\usepackage{soul}

\newcommand{\be}{\begin{equation}}
\newcommand{\ee}{\end{equation}}
\newcommand{\ba}{\begin{align}}
\newcommand{\ea}{\end{align}}
\newcommand{\sysb}{\left\{\begin{array}}
\newcommand{\syse}{\end{array}\right.}
\newcommand{\baa}{\begin{array}}
\newcommand{\eaa}{\end{array}}
\newcommand{\bs}{\begin{split}}
\newcommand{\es}{\end{split}}

\newcommand{\cosa}[1]{\cos \left(  #1 \right)}
\newcommand{\sina}[1]{\sin \left(  #1 \right)}

\newcommand{\comma}{\quad , \quad}
\newcommand{\mand}{\quad\text{ and }\quad}

\newcommand{\comm}[2]{\left[ #1, #2 \right]}

\newcommand{\trace}[1]{{\rm tr}\left\{ #1 \right\}}

\newcommand{\lan}{\left\langle}
\newcommand{\ran}{\right\rangle}
\newcommand{\abs}[1]{\left| #1 \right|}

\newcommand{\av}[1]{\lan #1 \ran}

\newcommand{\id}{\mathbb{1}}

\newcommand{\rme}[1]{{\rm{e}}^{#1}}

\newcommand{\ha}{\frac{1}{2}}
\newcommand{\Prob}{\mathbb{P}}

\newcommand{\lt}{\left(}
\newcommand{\rt}{\right)}

\usepackage{xr}
\usepackage{xcite}

\externalcitedocument[B-]{OQN_v12}

\begin{document}
\title{Quantum accelerated approach to the thermal state of classical spin systems with applications to pattern-retrieval in the Hopfield neural network}
\author{Eliana Fiorelli}
\affiliation{School of Physics and Astronomy, University of Nottingham, Nottingham, NG7 2RD, UK}
\affiliation{Centre for the Mathematics and Theoretical Physics of Quantum Non-equilibrium Systems, University of Nottingham, Nottingham NG7 2RD, UK}
\author{Pietro Rotondo}
\affiliation{School of Physics and Astronomy, University of Nottingham, Nottingham, NG7 2RD, UK}
\affiliation{Centre for the Mathematics and Theoretical Physics of Quantum Non-equilibrium Systems, University of Nottingham, Nottingham NG7 2RD, UK}
\author{Matteo Marcuzzi}
\affiliation{School of Physics and Astronomy, University of Nottingham, Nottingham, NG7 2RD, UK}
\affiliation{Centre for the Mathematics and Theoretical Physics of Quantum Non-equilibrium Systems, University of Nottingham, Nottingham NG7 2RD, UK}
\author{Juan P. Garrahan}
\affiliation{School of Physics and Astronomy, University of Nottingham, Nottingham, NG7 2RD, UK}
\affiliation{Centre for the Mathematics and Theoretical Physics of Quantum Non-equilibrium Systems, University of Nottingham, Nottingham NG7 2RD, UK}
\author{Igor Lesanovsky}
\affiliation{School of Physics and Astronomy, University of Nottingham, Nottingham, NG7 2RD, UK}
\affiliation{Centre for the Mathematics and Theoretical Physics of Quantum Non-equilibrium Systems, University of Nottingham, Nottingham NG7 2RD, UK}
\date{\today}
\begin{abstract}
We explore the question as to whether quantum effects can yield a speedup of the non-equilibrium evolution of spin systems towards a classical thermal state. In our approach we exploit the fact that the thermal state of a spin system can be mapped onto a node-free quantum state whose coefficients are given by thermal weights. This perspective permits the construction of a dissipative – yet quantum – dynamics which encodes in its stationary state the thermal state of the original problem. We show for the case of an all-to-all connected Ising spin model that an appropriate transformation of this dissipative dynamics allows to interpolate between a regime in which the order parameter obeys the classical equations of motion under Glauber dynamics, to a quantum regime with an accelerated approach to stationarity. We show that this effect enables in principle a speedup of pattern retrieval in a Hopfield neural network.
\end{abstract}
\maketitle
\emph{Introduction ---} A fundamental question that is currently triggering much attention in the quantum information, quantum many-body and computer science communities is whether quantum effects may lead to advantages in solving computational problems \cite{Nishimori:PRE:1999,Nishimori:JMP:2008,Lloyd:SIAM:2008,Troyer:NatPhys:2014,Baldassi:PNAS:2018,Biamonte:Nat:2017}. Several quantum algorithms have been proposed which can outperform their best classical counterparts, such as in the paradigmatic examples of integer factorization \cite{Shor:SIAM:1999} and database search problems \cite{Grover:PRL:1997}. Fluctuations due to quantum effects can moreover be employed to improve the performance of classical algorithms by opening ``tunnelling'' paths through high potential barriers that could otherwise trap a classical system in configurations potentially very different from the sought solution. This is the case, for instance, of quantum annealing \cite{Das2008} which seeks to find the state of minimum energy within the energy landscape of e.g. a highly-connected spin system with random couplings \cite{Farhi472,santoro2002theory,nishimori2017exponential,Denchev:PRX:2016}. More recently, a further paradigm emerged seeking to exploit the intrinsic open nature of quantum systems for quantum computing \cite{BreuerP:2002}. Its underlying idea is to encode the result of a computation in the stationary state of a suitably engineered \cite{diehl2008quantum, MULLER20121} quantum dissipative evolution of a many-body (spin) system \cite{Verstraete:NatPhys:2009}.

In this work we are interested in the question whether quantum effects in a purely dissipative dynamics can be advantageous for an accelerated approach to the thermal state of an interacting spin system. Analogously to the above-mentioned annealing or quantum computation protocols, this equilibrium state may encode the solution of a computational problem or the result of an optimization protocol. Our construction is based on a dissipative -- yet quantum -- generalization of a classical equilibrium Markov process. The corresponding dynamics has a pure stationary state which yields expectation values for classical observables that are identical to those of a thermal ensemble \cite{alicki1976}. We show that the stationary state is invariant under a set of unitary transformations which, however, affect the dynamics in a non-trivial way \cite{Olmos2014,Marcuzzi:JPA:2014}. This freedom permits a (quantum) speedup of the approach towards stationarity in comparison with the classical dynamics. We illustrate this in the case of a fully-connected Ising model and show that similar results hold for a Hopfield neural network (HNN) \cite{Hopfield:1982, Rotondo:JPA:2018}, which hints at the possibility of using quantum effects for the accelerated retrieval of patterns.

\begin{figure}[t]
\includegraphics[scale=0.8]{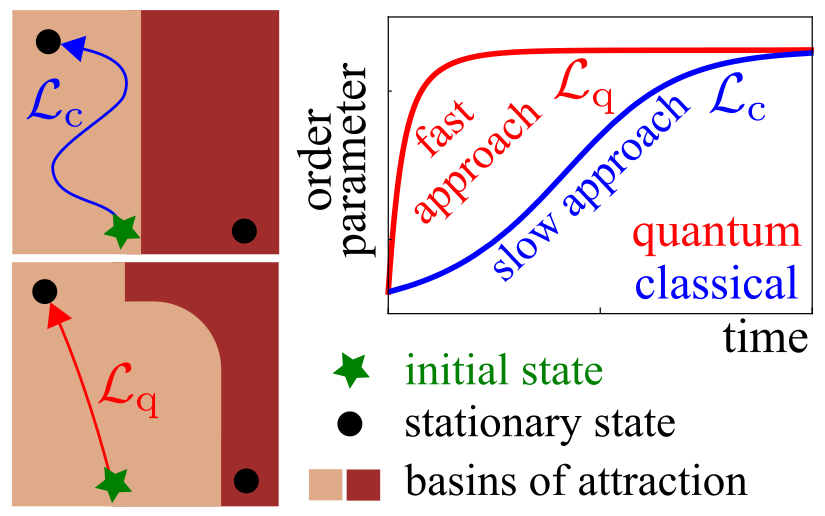}
\caption{\textbf{Quantum accelerated approach towards stationarity.} We consider a purely dissipative open quantum evolution which reproduces the stationary state properties of an equilibrium stochastic dynamics (with generator $\mathcal{L}_{\rm c}$). The quantum dynamics (generated by the operator $\mathcal{L}_{\rm q}$) will result generally in different dynamical timescales and may lead to an accelerated approach of the order parameter towards stationarity with respect to the classical evolution. The acceleration emerges from the unitary transformation of the dynamics that leaves the stationary states invariant but may change the structure of their basins of attraction.}
\label{fig1}
\end{figure}

\emph{Quantum generalization of classical stochastic processes ---} We focus for simplicity on equilibrium spin models. At the classical level, these are described by a set of configurations $\set{\vec{z}} = \set{z_1, z_2, \ldots, z_N}$ with $N$ the number of spins and $z_{i}=\pm 1$ some Ising spin variables. Each model is defined in terms of an energy (or ``cost'') function $E(\vec{z})$. In hard optimization problems \cite{Montanari:book}, these functions are typically defined so that their global minima correspond to the sought solutions in configuration space. Under any single-spin-flipping (or, more generally, ergodic) dynamics which satisfies detailed balance, the probability $P(\vec{z},t)$ of being in a given configuration $\vec{z}$ at time $t$ will approach, in the limit $t \to \infty$, the Gibbs distribution $P_{\rm eq}(\vec{z})=e^{-\beta E(\vec{z})}/Z(\beta)$, where $\beta=1/k_{B}T$ is the inverse temperature and $Z(\beta)= \sum_{\set{\vec{z}}}e^{-\beta E(\vec{z})} $ is the partition function.

To define a quantum generalization, we first promote the classical variables $z_i$ to quantum spins $\sigma_i^z$ and encode the corresponding configurations in quantum states $\ket{\vec{z}}$ such that $\sigma_i^z \ket{\vec{z}} = z_i \ket{\vec{z}}$. In the following, we shall refer to this as the ``classical basis'' and to observables diagonal in this basis as ``classical observables''. The state of the system is generically described by a density matrix $\rho$. We restrict for simplicity to Markovian and purely dissipative dynamics, so that in the Lindblad formalism \cite{Lindblad76, Breuer_P} $\rho$ evolves according to
\begin{equation}\label{eq.Lindblad}
\dot{\rho} = \mathcal{L} \rho = \sum_{j} \left( L_{j}\rho L_{j}^{\dagger}-\frac{1}{2}\left\lbrace L_{j}^{\dagger}L_{j}, \rho \right\rbrace \right).
\end{equation}
Here $\mathcal{L}$ denotes the generator of the time evolution and the operators $L_j$ are jump operators. In order to fix the form of the jump operators we define the pure state $\ket{\Psi_{\rm SS}} = \sum_{\set{\vec{z}}} \sqrt{P_{\rm eq}(\vec{z})} \ket{\vec{z}}$ and require it to be a \emph{dark state} of the dynamics, i.e., $L_j \ket{\Psi_{\rm SS}} = 0 \,\forall j$. This ensures that $\rho_{\rm SS} = \ket{\Psi_{\rm SS}} \bra{\Psi_{\rm SS}}$ is a stationary state of Eq.~\eqref{eq.Lindblad}, and that any expectation value of a classical observable on it corresponds to the (classical) thermal average $\braket{O_\mathrm{cl}} = \mathrm{Tr}(\rho_{\rm SS}O_\mathrm{cl})=\sum_{\set{\vec{z}}}e^{-\beta E(\vec{z})}\braket{\vec{z}|O_\mathrm{cl}|\vec{z}}/Z(\beta)$.
Hence, any classical property can be equivalently retrieved from this system, while quantum fluctuations can affect the typical timescales of the dynamics. We note that imposing $L_j \ket{\Psi_{\rm SS}} = 0 \,\forall j$ is reminiscent of the \emph{frustration-free} property typically associated to special (Rokhsar-Kivelson) systems \cite{Castelnovo:AnnPhys:2005}, as it ensures that the ``global'' stationary state property $\mathcal{L} \rho_{\rm SS} = 0$ is ``locally'' satisfied by each jump operator.

The dark-state property does not uniquely fix the jump operators. For instance, if from a given set of $L_j$ one were to construct a second one $L_j' = u_j L_j$ with $u_j$ a set of unitary transformations ($u_j^\dag u_j = \id$), the dynamics would still have $\rho_{\rm SS}$ as a stationary state (since $L_j' \ket{\Psi_{\rm SS}} = 0 \,\,\forall j$). Note that it is the ``frustration-free'' property of the dark state that allows one to choose different unitary operators $u_j$ for different $j$. As sketched in Fig.~\ref{fig1}, this freedom allows to construct and explore different dynamics, all sharing the same stationary state(s). Generically, the typical timescales of the dynamics will change for different $u_j$, yielding in some cases a faster, in others a slower approach to stationarity.

In the following we construct explicitly a set of jump operators. To this end we define $\vec{z}_{\set{i}}$ as the list of Ising variables excluding the $i$-th one. Furthermore, we denote by $D_i (\vec{z}_{\set{i}})= [E(\vec{z}_{\set{i}}, z_i = -1) - E(\vec{z}_{\set{i}}), z_i = +1)]/2$ half the energy cost for flipping the $i$-th spin down, leaving the configuration of the remaining spins fixed. From $D_i$ we define an operator $\Delta E_i$ obtained by taking the functional form of $D_i$ and replacing every $z_j$ ($j \neq i$) with the corresponding Pauli matrix $\sigma_j^z$. For example, for $D_i = (1/N) \sum_{j\neq i} z_j$ one would get $\Delta E_i = (1/N) \sum_{j\neq i} \sigma^z_j$. The $N$ jump operators (one per site) are then constructed according to
\begin{equation}\label{eq.jumpop}
L_{i}=\alpha^{-}_{i}n_{i}-\alpha^{+}_{i}\sigma_{i}^{+}, \quad\mathrm{with} \quad \alpha_i^\pm = \frac{\rme{\pm \frac{\beta}{2} \Delta E_i}}{\left[  2\cosh \lt \beta \Delta E_i \rt \right]^\ha}.
\end{equation}
Here $\sigma_{i}^{\pm}=(\sigma_{i}^{x}\pm i \sigma_{i}^{y})/2$ are the Pauli raising and lowering operators, $n_{i}= (\sigma_{i}^{z}+\mathbb{1}_{2})/2 $, and the dependence of $\Delta E_{i}$ on all spins but the $i$-th one is implicit. This particular form of the jump operators is convenient, because it will allow us, in the examples discussed further below, to directly relate the purely-dissipative quantum dynamics to the corresponding classical Glauber dynamics with energy function $E (\vec{z})$.

To conclude the construction of the jump operators we use the freedom given by the unitaries $u_j$. For the sake of simplicity we choose local unitaries parameterized by the two angles $\theta$ and $\phi$: $u_j (\theta, \phi)=e^{i\phi \frac{\sigma_{i}^{z}}{2}}e^{i\theta \sigma_{i}^{y}}e^{-i \phi \frac{\sigma_{i}^{z}}{2}}$.

\emph{Fully-connected quadratic models ---} To study the dependence of the relaxation timescales under the angles $(\theta, \phi)$, we focus here on fully-connected models $E(\vec{z}) = -\sum_{i,j} J_{ij} z_i z_j$, with $J_{ij}$ being a symmetric real $N \times N$ matrix. This will allow us to study the dynamics in terms of semi-classical collective variables. The equations of motion for the local operators $\sigma_i^{x,y,z}$ generated by the dynamics \eqref{eq.Lindblad} read (see \cite{SM})
\begin{eqnarray}
 \dot{\sigma}_{i}^{x} &=&-A_{i}^{x} -\frac{\beta}{2}\sum_{k \neq i } J_{k i} \mathrm{sech}(\beta \Delta E_{k}) \sigma_{k}^{y} \sigma_{i}^{y}+ \frac{ \mathrm{sech}(\beta \Delta E_{i})}{2}, \nonumber\\
 \dot{\sigma}_{i}^{y} &=&-A_{i}^{y} +\frac{\beta}{2}\sum_{k \neq i } J_{k i} \mathrm{sech}(\beta \Delta E_{k}) \sigma_{k}^{y}\sigma_{i}^{x},\nonumber\\
 \dot{\sigma}_{i}^{z} &=& -A_{i}^{z}+ \frac{1}{2}\tanh(\beta \Delta E_{i}).\label{eq.spin}
\end{eqnarray}
Here $A_{i}^{\alpha}=\lbrace f_{\alpha}\left[\mathrm{sech}(\beta \Delta E_{i}) \sigma_{i}^{x}+\tanh(\beta \Delta E_{i})\sigma_{i}^{z} -1\right]+\sigma_{i}^{\alpha}\rbrace/2$ ($\alpha=x,y,z$), which depends on the angles $(\theta,\phi)$ through the functions $f_{x}(\theta,\phi)=-\sin(2\theta)\cos(\phi)$, $f_{y}(\theta,\phi)=\sin(2\theta)\sin(\phi)$, $f_{z}(\theta)= \cos(2\theta)$. Note that for $\theta = \pi/4$ [i.e., $f_z(\pi/4) = 0$] the last equation is defined entirely in terms of combinations of $\sigma^z$ matrices. Further below, we shall use this choice (together with $\phi = \pi/2$) as a classical reference case, as it will yield, for the $z$-component of spin operator, the same dynamics one would derive from a purely-classical Glauber dynamics.

\emph{Fully-connected Ising model ---} As a first example we consider the fully-connected Ising model, i.e., we choose $J_{ij} = 1/N$. We construct a set of semiclassical collective variables that allow to reduce the problem to a system of $3N$ coupled ordinary differential equations: $s^{\alpha}=(1/N)\sum_{i=1}^{N}\sigma_{i}^{\alpha}$ ($\alpha= x,y,z$). Since their commutator $\comm{s^\alpha}{s^\beta} = 2i\epsilon_{\alpha \beta \gamma} s^\gamma / N $ vanishes in the thermodynamic limit $N\to \infty$, we can effectively replace them with their expectation values $s^\alpha \approx \av{s^\alpha} \equiv m^\alpha$, leading to
\begin{eqnarray}
 \dot{m}^{x}&=&-A_{x} + \frac{1}{2}\mathrm{sech}(\beta m^{z}) -\frac{\beta}{2} \mathrm{sech}(\beta m^{z}) (m^{y})^{2},\nonumber\\
 \dot{m}^{y}&=&-A_{y} + \frac{\beta}{2} \mathrm{sech}(\beta m^{z}) m^{y} m^{x},\nonumber\\
 \dot{m}^{z}&=&-A_{z} + \frac{1}{2}\tanh(\beta m^{z}). \label{e.ferromagnetic}
\end{eqnarray}
Here $A_{\alpha} = \lbrace f_{\alpha}\left[ m^{x}\mathrm{sech}(\beta m^{z}) + m^{z}\tanh(\beta m^{z}) -1 \right]+m^{\alpha}\rbrace/2 $ ($\alpha=x,y,z$). The choice $\theta = \pi/4$ decouples the third equation from the others, implying that the dynamics of $m^z$ proceeds independently from the one of $m^{x,y}$. Furthermore, the equation for $m^z$ is equivalent to the mean-field evolution equation of the order parameter of a classical Ising model under a continuous-time Glauber dynamics. We thereby consider this angle $\theta = \pi/4$ our \emph{classical} reference point (the specific choice of $\phi$ will not affect the evolution of $m^z$).

The stationary state structure can be extracted by setting the l.h.s.~of Eqs.~\eqref{e.ferromagnetic} to zero: for $\beta \leq 1$ (high temperature) the equations admit a single (``paramagnetic'') solution $m^z = m^y = 0$, $m^x = 1$. For $\beta < 1$ (low temperature) the paramagnetic solution becomes unstable and two stable ``ferromagnetic'' ($m^z \neq 0$) solutions appear obeying $m^z = \tanh (\beta m^z)$.

\begin{figure}[t]
\includegraphics[scale=0.45]{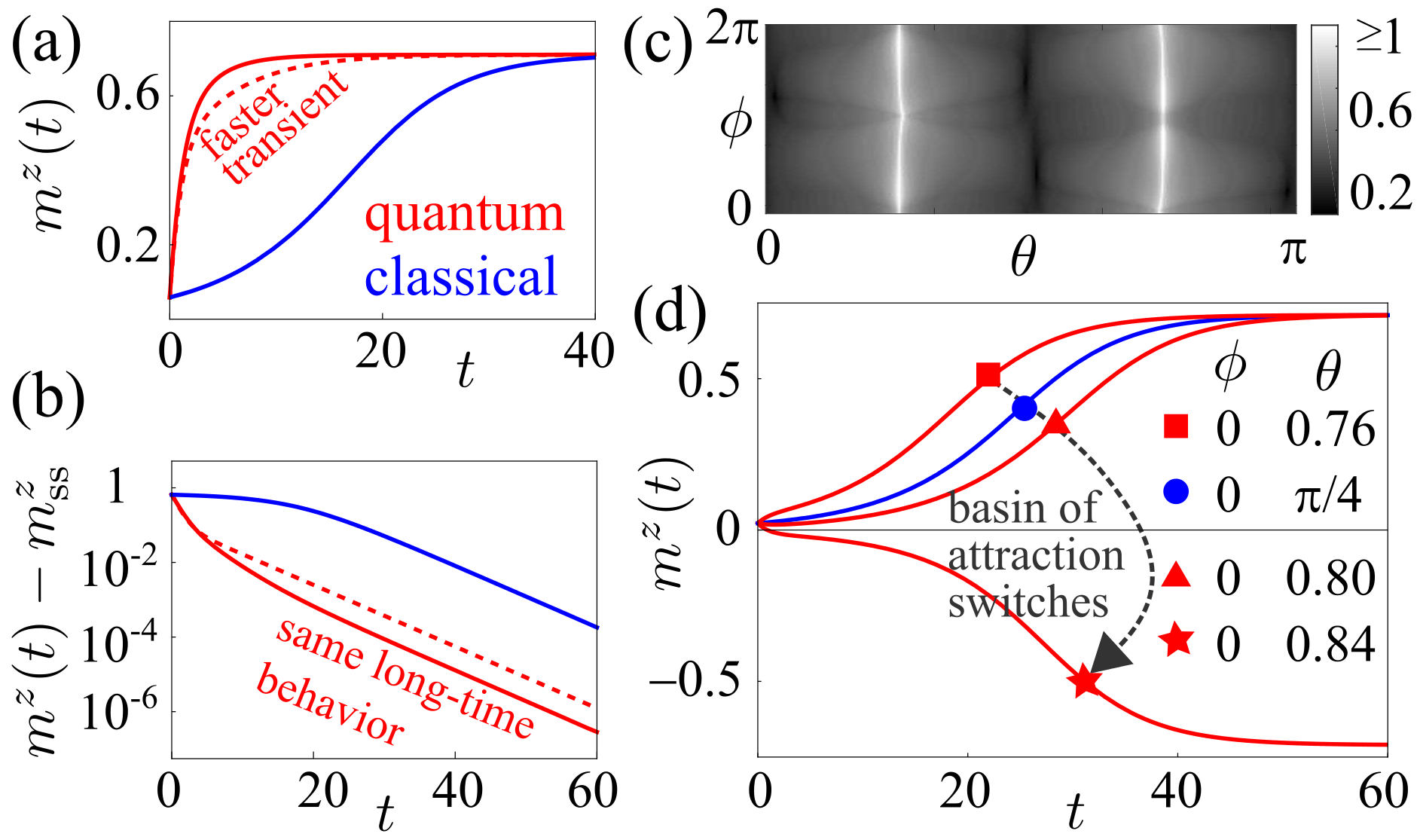}
\caption{\textbf{Comparison between the ``classical'' and ``quantum'' dynamics of a fully-connected Ising model.} (a) Magnetization $m^{z}$ as a function of time $t$. (b) Logarithmic plot of $m^{z}(t) - m^z(t \to \infty)$ displaying the exponential approach towards the stationary value of the magnetization. The equal slope of all curves highlights that the asymptotic rate is the same for all choices of the angles ($\theta, \phi$). (c) Density plot of the ratio $\bar{t}/\bar{t}_\mathrm{c}$ in the $(\theta,\phi)$-plane. Here $\bar{t}$ is the relaxation time at given $\theta$, $\phi$-values and $\bar{t}_\mathrm{c}$ is the relaxation time in the classical limit (at $\theta = \pi/4$). (d) Magnetization $m^{z}$ for selected angles [see panel (c)]. In panels (a) and (b) blue denotes the classical case $\theta = \pi/4$; red lines correspond to the choices $(\theta,\phi)=(3,0)$ (dashed) and $(\theta,\phi)=(0.3, 4)$ (solid). In all panels, $\beta^{-1}=0.8$ and the initial conditions are randomly chosen such that $|m^{\alpha}(0)| \ll 1$ ($\alpha=x,y,z$).}
\label{fig2}
\end{figure}

In Fig.~\ref{fig2}(a) we show the ``quantum'' (red lines) and ``classical'' (blue line) evolution of the order parameter $m^z (t)$ in the ferromagnetic phase for a specific choice of $\theta$ and $\phi$ and fixed initial conditions. The observed acceleration away from the classical limit arises in the early stages of the dynamics, whereas the long-time, asymptotic behavior is exponential with the same rate for both curves [panel (b)]. The onset of this exponential decay is thus shifted in time, leading to a speedup.

To quantify this accelerated approach to stationarity, we define the relaxation time $\bar{t}$ as the time it takes the order parameter to approach its stationary value within the threshold $\epsilon = 10^{-3}$ (see \cite{SM} for more details). The choice of this threshold is arbitrary; however, for most values of $\theta$ and $\phi$ the asymptotic approach to stationarity is exponential with the \emph{same} rate for both quantum and classical dynamics, as in Fig.~\ref{fig2}(b). Hence, the specific value of $\epsilon$ is not relevant to determine the presence of a speedup. Some small regions where the quantum dynamics is asymptotically slower than the classical one exist, but they are included within regions where the early-time dynamics is slowed down as well.
In Fig.~\ref{fig2}(c) we show $\bar{t}/\bar{t}_\mathrm{c}$ --- which is the relaxation time normalized by the classical relaxation time $\bar{t}_\mathrm{c}$ (at $\theta = \pi/4$ and $\phi = \pi/2$) --- in the $(\theta,\phi)$-plane for a given choice of very small initial conditions. The plot shows large regions of accelerated relaxation separated by two narrow strips, close to the ``classical'' regime, where the quantum dynamics experiences instead a slowdown. The shape of these strips shows only a weak dependence on the initial conditions.



This robustness is a consequence of the fact that, while the stationary states are independent of the angles $\theta$ and $\phi$, their respective basins of attraction change (see sketch in Fig.~\ref{fig1}). As can be gleaned from Eqs.~\eqref{e.ferromagnetic}, $m^z = 0$ identifies an invariant subspace under the classical dynamics (i.e., $m^z = 0 \Rightarrow \dot{m}^z = 0$). Hence, by choosing the initial conditions $\abs{m^\alpha} \ll 1$ the initial dynamics in the $z$-direction will necessarily be slow (due to the almost vanishing derivative). By changing the angles (and thereby the shape of the attraction basins), the same initial condition will generically not be close to such an invariant manifold and the quantum dynamics will start faster, leading to the observed accelerated early-time dynamics. Indeed, the instances in which it is slower [bright regions in panel (c)] correspond to regions where the dynamics swaps from reaching one ferromagnetic stationary state to its opposite [panel (d)]. Hence, for most choices of $\theta$ and $\phi$ the quantum dynamics will be faster than the classical one independently of the initial conditions as long as $|m^{\alpha}(0)| \ll 1$.


\emph{Application to the Hopfield neural network ---} The HNN is a fundamental model for an associative memory capable of storing a set of $p$ spin configurations $\set{\xi_1^{\mu}, \ldots , \xi_N^{\mu}}$ ($\mu=1,...,p$). These are referred to as \emph{patterns}, where each component takes the values $\xi_i^{\mu} = \pm 1$ ($i=1,...,N$). Hereafter, we shall adopt the vector notation for the $p$-dimensional pattern space, e.g., $\vec{\xi_i} = (\xi_i^1, \ldots , \xi_i^p)^T$. For large $N$, the pattern components are assumed to have a random structure symmetrically distributed between $\pm 1$, so that $(1/N) \sum_i \xi_i^\mu \approx 0$ and $(1/N) \sum_i \xi_i^\mu \xi_i^\nu \approx \delta_{\mu \nu}$. More specifically, the $\xi_i^\mu$ are described as a set of independent, identically-distributed random variables with distribution $\Prob\lt \xi_i^\mu = \pm 1 \rt = \tfrac{1}{2}$. For the HNN, the coupling matrix reads $J_{ij}=(1/N)\sum_{\mu=1}^{p}\xi_{i}^{\mu}\xi_{j}^{\mu}$. The corresponding energy function is minimized by choosing configurations $z_i = \xi_i^{\mu} \, \forall i$ for any fixed $\mu$. Hence, a stochastic Glauber dynamics may have the patterns as stationary states, as long as the number of patterns obeys $p/N \ll 1$ \citep{AmitGS:1985a,AmitGS:1985b}. More generally, at finite temperature two phases emerge: for $\beta < 1$ the system is in a paramagnetic phase where the typical configurations have no extensive overlap with any of the patterns [$\lim_{N \to \infty} (1/N)\sum_i \xi_i^\mu \av{z_i} = 0 \,\forall \mu$]. Instead, $\beta > 1$ identifies a ``retrieval'' phase where the system picks one of the patterns (say, the $\nu$-th) and acquires a non-vanishing overlap with it [$\lim_{N \to \infty} (1/N)\sum_i \xi_i^\nu \av{z_i} \neq 0$].

In the corresponding quantum model, this overlap is generalized to $m_\mu^\alpha = (1/N)\sum_{i=1}^{N} \xi^{\mu}_{i} \av{\sigma_{i}^{\alpha}}$ with $\av{\cdot} = \trace{(\cdot) \rho(t)}$ representing the quantum expectation value at time $t$. However, in contrast to the fully-connected Ising model, the equations of motion do not straight-forwardly close in the collective variables (overlaps) $m_\mu^\alpha$ and further approximations are required: since the stationary state should yield the same expectation values as a classical HNN, it is natural to think that, at least not too far from the stationary points, the dynamics will be mostly determined by the properties of the overlaps $m_\mu^\alpha$. We thus perform the approximation $\av{\sigma_i^\alpha} \approx \xi_i^\mu m_\mu^\alpha$ (consistency checks are discussed in \cite{SM}). Moreover, we exploit the self-averaging property, i.e.~that for large $N$ we can perform the substitution $(1/N) \sum_{i}F(\vec{\xi}_{i})\rightarrow \overline{F(\vec{\xi})}$, where $\overline{(\cdot)}$ denotes the average over the disorder. This reduces the equations of motion to \cite{SM}
\begin{eqnarray}\label{e.hopfieldgen}
 \dot{m}_{\mu}^{x,y} &=& -\bar{A}_{\mu}^{x,y}(\beta), \nonumber\\
 \dot{m}_{\mu}^{z} &=& -\bar{A}_{\mu}^{z}(\beta) + \frac{1}{2}  \overline{\xi^{\mu}\tanh{\beta \vec{\xi} \cdot \vec{m}^{z}}},
\end{eqnarray}
where $ \bar{A}_{\mu}^{\alpha}= \left[f_{\alpha} m_{\mu}^{x} \overline{ \mathrm{sech}(\beta \vec{\xi} \cdot \vec{m}^{z})}+m_{\mu}^{\alpha}\right]/2$ ($\alpha = x,y,z$). These equations close in the $3p$ variables $m_\mu^{x,y,z}$ and can be solved numerically, the averages being calculated according to their definition $\overline{F(\vec{\xi})} = \tfrac{1}{2^p} \sum_{\xi^1 = \pm 1} \ldots \sum_{\xi^p = \pm 1} F(\vec{\xi})$.

\begin{figure}[t]
\includegraphics[scale=0.45]{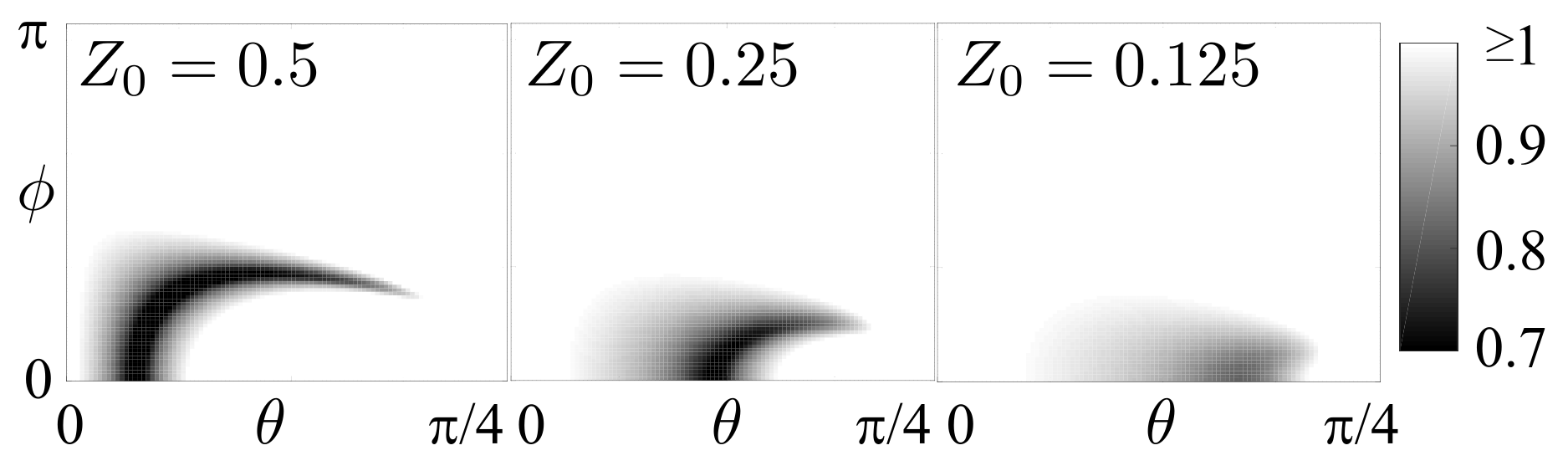}
\caption{\textbf{Comparison between the ``classical'' and ``quantum'' dynamics of a HNN.} Density plot of the ratio $\bar{t}/\bar{t}_\mathrm{c}$, averaged over $200$ realizations (for protocol see main text). The initial conditions are fixed such that $|m_{\mu}^{z}(0)|$ is uniformly distributed in the interval $[-0.01, 0.01]$ $\forall \mu=1,...,p-1$, whereas $m_p^z (0) \in [Z_0-0.01,Z_0+0.01]$ with $Z_0 = 0.5$, $0.25$ and $0.125$. The remaining components $m^x (0)$ and $m^y(0)$ are randomly chosen in $[-0.01, 0.01]$. For the ``quantum'' case the rotation $m^z(0) \to m^x (0)$, $m^x(0) \to - m^z(0)$ is performed before running the dynamics. The remaining parameters, common to all panels, are $p=3$ (number of patterns), $\beta^{-1} = 0.8$. These plots display one eighth of the entire $(\theta,\phi)$-plane [corresponding to the bottom-left corner of Fig.~\ref{fig2}(c)], but the remainder can be reconstructed by symmetry arguments via the transformations $\phi \to 2\pi - \phi$, $\theta \to \pi/2 - \theta$ and $(\theta, \phi) \to (\pi - \theta, \phi + \pi)$ (see \cite{SM} for more details).}
\label{fig3}
\end{figure}

As in the fully-connected Ising model, the choice $\theta = \pi/4$ decouples the equations for the $z$-component and makes them equivalent to the classical (mean-field) dynamics.
Unlike in the previous case, however, the order parameters [the overlaps $m^\alpha (t)$] are not necessarily known (one would have to know the patterns $\xi^\mu$ to derive them from the spin configurations) and therefore an averaging over initial conditions is required.


We focus on initial conditions representing ``corrupted memories'', i.e., we pick the initial (random) overlaps $m^{z}_\mu (0)$ uniformly over a small interval $[-0.01, 0.01]$ with the exception of a single one (without loss of generality, we fix $\mu = p$) $m^z_p (0) = Z_0 \pm 0.01$, which is instead centered around a non-vanishing value $Z_0 \neq 0$. This mimics a case where a classical memory encodes a portion of a given pattern and the task of the neural network is to reconstruct the remainder. 



Notably, merely changing $\theta$ and $\phi$ does not lead to any speedup as in the Ising case. To find a regime where quantum acceleration occurs, a possible protocol is the following: for each random choice of the initial conditions a spin rotation is performed which maps the overlaps' $x$-components in $z$-components and vice versa [$m^{x}_{\mu}(0) \rightarrow - m^{z}_{\mu}(0)$, $m^y_\mu (0) \rightarrow m^y_\mu (0)$, $m^{z}_{\mu}(0) \rightarrow m^{x}_{\mu}(0)$]. Afterwards the quantum dynamics is switched on. To quantify the speedup the timescale $\bar{t}$ is measured analogous to the Ising case and compared with the classical value $\bar{t}_c$. The ratios $\bar{t} / \bar{t}_c$ corresponding to different initial conditions are subsequently averaged.
%
%
The results are shown in Fig.~\ref{fig3}, where this ratio is displayed in the ($\theta$,$\phi$)-plane for three choices of $Z_0$. We can identify choices of the angles for which an accelerated pattern retrieval is indeed achieved. In particular, the gain is more significant for less corrupted initial memories, i.e. larger $Z_0$.

\emph{Conclusions ---} We investigated how quantum effects can accelerate the approach towards stationarity of a classical stochastic system. Our results suggest that indeed complementary methods to the established quantum annealing techniques \cite{Nishimori:PRE:1999,Nishimori:JMP:2008,Lloyd:SIAM:2008,Troyer:NatPhys:2014,Baldassi:PNAS:2018,Denchev:PRX:2016} may be exploited to enhance pattern retrieval in neural networks. In the future it will be interesting to extend this idea to more complex settings, for example spin systems with disorder that realize specific instances of NP-hard problems, such as the Sherrington-Kirkpatrick model \cite{Mezard:book} or the spin glass phase of the HNN \citep{AmitGS:1985b}. Furthermore, it would be interesting to study experimental implementations of the proposed quantum dynamics based on open multimodal cavity arrays, which have been proposed as emulators for quantum neural networks and glassy systems \cite{Goldbart:PRL:2011,Strack:PRL:2011,Rotondo:PRL:2015,Rotondo:PRB:2015,Lev:PRX:2018}.

\emph{Acknowledgments ---} We acknowledge discussions with M. M\"uller. The research leading to these results has received funding from the European Research Council  under  the  European  Unions  Seventh  Framework  Programme  (FP/2007-2013)/ERC  Grant  Agreement  No. 335266  (ESCQUMA). P.R. acknowledges funding by the European Union through the H2020 - MCIF No. $766442$. I.L. gratefully acknowledges funding  through  the  Royal  Society  Wolfson  Research Merit Award.

\begin{widetext}
\section{Supplemental material for: Quantum accelerated approach to the thermal state of classical spin systems with applications to pattern-retrieval in the Hopfield neural network}
\section{Derivation of Eqs. (\ref{B-eq.spin}) of the main text}

We firstly consider the equations of motion generated by the Markovian and purely dissipative dynamics defined in Eq.(\ref{B-eq.Lindblad}) with the jumps operators (\ref{B-eq.jumpop}), only secondly applying the unitary transformation $u_{i}(\theta, \phi)$ obtaining the Eqs.~(\ref{B-eq.spin}). Here, any operator $O$ evolves according to the adjoint Lindblad equation, $\dot{O}=\sum_{k}L_{k}^{\dagger}O L_{k}-\frac{1}{2} \lbrace L^{\dagger}_{k}L_{k},O \rbrace$. 

The equations of motion for the operators $\sigma^{\gamma}_{i}$, $\gamma=x,y,z$ read

\begin{equation}\label{sm.spin}
\dot{\sigma}_{i}^{\gamma} = \sum_{k}L_{k}^{\dagger}\sigma_{i}^{\gamma}L_{k}-\frac{1}{2}\lbrace L_{k}^{\dagger}L_{k}, \sigma_{i}^{\gamma} \rbrace.
\end{equation}
Let us first specialize (\ref{sm.spin}) for the $\sigma^{z}_{i}$ operator. As it is $[\alpha^{\pm}_{k}, \sigma_{i}^{z} ] = 0$ $\forall k,i$, the terms arising from the sum over 
$k \neq i$ vanish, and Eq.(\ref{sm.spin}) reads
\begin{equation}\label{e.sigmaz_fin}
\begin{split}
 \dot{\sigma}_{i}^{z}= & (\alpha^{-}_{i}n_{i}-\alpha^{+}_{i}\sigma_{i}^{-})\sigma_{i}^{z}(\alpha^{-}_{i}n_{i}-\alpha^{+}_{i}\sigma_{i}^{+})-\frac{1}{2} \left\lbrace (\alpha^{-}_{i}n_{i}-\alpha^{+}_{i}\sigma_{i}^{-})(\alpha^{-}_{i}n_{i}-\alpha^{+}_{i}\sigma_{i}^{+}),\sigma_{i}^{z} \right\rbrace=\\
 = & (\alpha^{-}_{i})^{2}n_{i}+(\alpha^{+}_{i})^{2}\sigma_{i}^{-}\sigma_{i}^{+}- \alpha^{-}_{i}\alpha^{+}_{i}\sigma_{i}^{+}-\alpha^{-}_{i}\alpha^{+}_{i}\sigma_{i}^{-} - \frac{1}{2}(\alpha^{-}_{i})^{2}n_{i} +\frac{1}{2}(\alpha^{+}_{i})^{2}\sigma_{i}^{-}\sigma_{i}^{+} + \\
& - \frac{1}{2}\alpha^{-}_{i}\alpha^{+}_{i}\sigma_{i}^{+} +\frac{1}{2}\alpha^{-}_{i}\alpha^{+}_{i}\sigma_{i}^{-} - \frac{1}{2}(\alpha^{-}_{i})^{2}n_{i} + \frac{1}{2}(\alpha^{+}_{i})^{2}\sigma_{i}^{-}\sigma_{i}^{+} +\frac{1}{2}\alpha^{-}_{i}\alpha^{+}_{i}\sigma_{i}^{+} \\
& - \frac{1}{2}\alpha^{-}_{i}\alpha^{+}_{i}\sigma_{i}^{-} =  -\alpha^{-}_{i}\alpha^{+}_{i}\sigma_{i}^{x} + (\alpha^{+}_{i})^{2}(1-\sigma_{i}^{z}) =\\
= & - \frac{1}{2}\mathrm{sech}(\beta \Delta E_{i})\sigma_{i}^{x} + \frac{1}{2}\left[1+ \tanh( \beta \Delta E_{i})\right](1-\sigma_{i}^{z}),
\end{split}
\end{equation}
where it has been used that $n_{i}^{2}=n_{i}$, $n_{i}\sigma_{i}^{+}=\sigma_{i}^{+}$, $\sigma_{i}^{-}n_{i}=\sigma_{i}^{-}$. We can now consider the equation of motion of the operator $\sigma_{i}^{+}$, which reads
\begin{equation}\label{e.sigma+}
\begin{split}
& \dot{\sigma}_{i}^{+} = \sum_{k} \lbrace (\alpha^{-}_{k}n_{k}-\alpha^{+}_{k}\sigma_{i}^{-})\sigma_{i}^{+}(\alpha^{-}_{k}n_{k}-\alpha^{+}_{k}\sigma_{i}^{+})- \\
& \frac{1}{2} \left[ (\alpha^{-}_{k}n_{k}-\alpha^{+}_{k}\sigma_{i}^{-})(\alpha^{-}_{k}n_{k}-\alpha^{+}_{k}\sigma_{i}^{+})\sigma_{i}^{+} + \sigma_{i}^{+}(\alpha^{-}_{k}n_{k}-\alpha^{+}_{k}\sigma_{i}^{-})(\alpha^{-}_{k}n_{k}-\alpha^{+}_{k}\sigma_{i}^{+})\right] \rbrace.
\end{split}
\end{equation}
The term arising from the sum over $k=i$ is
\begin{equation}
\begin{split}
& (\alpha^{-}_{i})^{2}\sigma_{i}^{+}n_{i} + (\alpha^{+}_{i})^{2}\frac{(1-\sigma_{i}^{z})}{2}\sigma_{i}^{+} -\alpha^{-}_{i} \alpha^{+}_{i}\sigma_{i}^{+}\sigma_{i}^{+} - \alpha^{-}_{i} \alpha^{+}_{i}\sigma_{i}^{-}\sigma_{i}^{+}n_{i} - \frac{1}{2} (\alpha^{-}_{i})^{2}\sigma_{i}^{+} - \\
& - \frac{1}{2}(\alpha^{+}_{i})^{2}\frac{(1-\sigma_{i}^{z})}{2}\sigma_{i}^{+} + \frac{1}{2}\alpha^{-}_{i} \alpha^{+}_{i}\sigma_{i}^{+}\sigma_{i}^{+} + \frac{1}{2}\alpha^{-}_{i} \alpha^{+}_{i}\frac{(1-\sigma_{i}^{z})}{2} - \frac{1}{2}(\alpha^{-}_{i})^{2}\sigma_{i}^{+}n_{i} \\
& -\frac{1}{2}(\alpha^{+}_{i})^{2}\frac{(1+\sigma_{i}^{z})}{2}\sigma_{i}^{+} + \frac{1}{2}\alpha^{-}_{i} \alpha^{+}_{i}\sigma_{i}^{+}\sigma_{i}^{+} + \frac{1}{2} \alpha^{-}_{i} \alpha^{+}_{i}\frac{(1+\sigma_{i}^{z})}{2} = \\
& = - \frac{(\alpha^{-}_{i})^{2}+(\alpha^{+}_{i})^{2}}{2}\sigma_{i}^{+} + \frac{\alpha^{-}_{i}\alpha^{+}_{i}}{2} =- \frac{\sigma_{i}^{+} }{2}+ \frac{\alpha^{-}_{i}\alpha^{+}_{i}}{2}.
\end{split}
\end{equation}
Before evaluating the part of Eq.(\ref{e.sigma+}) coming from the sum over $k \neq i$, as it is $[\alpha^{\pm}_{k}, \sigma_{i}^{+} ]\neq 0$ for $i \neq k$, it is useful to consider the following expression
\begin{equation}
\begin{split}
 \sigma_{i}^{+}\alpha^{+}_{k} = & \sigma_{i}^{+} \frac{e^{\frac{\beta}{2}\sum_{j}J_{kj}\sigma_{j}^{z}}}{\sqrt{2 \cosh( \beta\sum_{j}J_{kj}\sigma_{j}^{z})}} = \frac{ e^{\frac{\beta}{2}\sum_{j\neq i}J_{kj}\sigma_{j}^{z}}e^{-\frac{\beta}{2}J_{ki}}}{\sqrt{2 \cosh( \beta\sum_{j \neq i}J_{kj}\sigma_{j}^{z} - \beta J_{ki})}}\sigma_{i}^{+} = \\
= & \frac{e^{\frac{\beta}{2}\sum_{j}J_{kj}\sigma_{j}^{z}} e^{-\frac{\beta}{2}J_{ki}} e^{-\frac{\beta}{2}J_{ki}\sigma_{i}^{z}}}{\sqrt{2 \cosh( \beta\sum_{j \neq i}J_{kj}\sigma_{j}^{z} - \beta J_{ki})}}\sigma_{i}^{+} = \\
= & \underbrace{\frac{e^{\frac{\beta}{2}\sum_{j}J_{kj}\sigma_{j}^{z}}}{\sqrt{2 \cosh( \beta\sum_{j}J_{kj}\sigma_{j}^{z})}}}_{= \alpha^{+}_{k}} \underbrace{ \left( \frac{\cosh( \beta\sum_{j}J_{kj}\sigma_{j}^{z})}{\cosh( \beta\sum_{j \neq i}J_{kj}\sigma_{j}^{z} - \beta J_{ki})} \right)^{\frac{1}{2}}}_{ \equiv f_{ki}} e^{-\beta J_{ki}} \sigma_{i}^{+} = \\
= & \alpha^{+}_{k}f_{ki}e^{-\beta J_{ki}} \sigma_{i}^{+}, \\ 
\end{split}
\end{equation}
which, after having defined $\alpha^{+}_{k}f_{ki} e^{-\beta J_{ki}} \equiv \alpha^{+}_{ki}$, can be written as

\begin{equation}
\sigma_{i}^{+}\alpha^{+}_{k} = \alpha^{+}_{ki}\sigma_{i}^{+}.
\end{equation}
Similarly, it is
\begin{equation}
\begin{split}
& \sigma_{i}^{+}(\alpha^{+}_{k})^{2} = (\alpha^{+}_{ki})^{2}\sigma_{i}^{+}, \\
& \sigma_{i}^{+}\alpha^{-}_{k} = \alpha^{-}_{ki}\sigma_{i}^{+}, \\
& \sigma_{i}^{+}(\alpha^{-}_{k})^{2} = (\alpha^{-}_{ki})^{2}\sigma_{i}^{+},
\end{split}
\end{equation}
where $\alpha^{-}_{ki} \equiv \alpha^{-}_{k} f_{ki}e^{\beta J_{ki}}$. Let us now go ahead in evaluating the last part of Eq.(\ref{e.sigma+}), which reads
\begin{equation}
\begin{split}
& \sum_{k \neq i} \left\lbrace  \alpha^{-}_{k}n_{k}\sigma_{i}^{+}\alpha^{-}_{k}n_{k} + \alpha^{+}_{k}\sigma_{k}^{-}\sigma_{i}^{+}\alpha^{+}_{k} \sigma_{k}^{+} - \alpha^{-}_{k}n_{k}\sigma_{i}^{+}\alpha^{+}_{k} \sigma_{k}^{+} - \alpha^{+}_{k}\sigma_{k}^{-}\sigma_{i}^{+}\alpha^{-}_{k}n_{k} - \frac{1}{2} \left[ (\alpha^{-}_{k})^{2}n_{k}\sigma_{i}^{+} + \right. \right. \\
& \left. \left.  + (\alpha^{+}_{k})^{2}\sigma_{k}^{-}\sigma_{k}^{+}\sigma_{i}^{+} - \alpha^{-}_{k}\alpha^{+}_{k}\sigma_{k}^{x}\sigma_{i}^{+} + \sigma_{i}^{+}(\alpha^{-}_{k})^{2}n_{k} +  \sigma_{i}^{+}(\alpha^{+}_{k})^{2}\sigma_{k}^{-}\sigma_{k}^{+} - \sigma_{i}^{+}\alpha^{-}_{k}\alpha^{+}_{k}\sigma_{k}^{x} \right] \right\rbrace = \\
& = \sum_{k \neq i} \left( \alpha^{-}_{k}\alpha^{-}_{ki}n_{k} - \alpha^{+}_{k}\alpha^{-}_{ki}\sigma_{k}^{-} - \frac{1}{2}(\alpha^{-}_{ki})^{2}n_{k} + \alpha^{+}_{k}\alpha^{+}_{ki}\sigma_{k}^{-}\sigma_{k}^{+} - \alpha^{-}_{k}\alpha^{+}_{ki}\sigma_{k}^{+} - \right. \\
& \left. - \frac{1}{2}(\alpha^{+}_{ki})^{2}\sigma_{k}^{-}\sigma_{k}^{+} + \frac{1}{2}\alpha^{-}_{ki}\alpha^{+}_{ki}\sigma_{k}^{x} - \frac{1}{2}(\alpha^{-}_{k})^{2}n_{k} - \frac{1}{2}(\alpha^{+}_{k})^{2}\sigma_{k}^{-}\sigma_{k}^{+} + \frac{1}{2} \alpha^{-}_{k}\alpha^{+}_{k}\sigma_{k}^{x} \right) \sigma_{i}^{+} = \\
& = \sum_{k \neq i}  \frac{1}{2} \left[ -\left( \alpha^{-}_{k} - \alpha^{-}_{ki} \right)^{2}n_{k}  - ( \alpha^{+}_{k} - \alpha^{+}_{ki} )^{2}\sigma_{k}^{-}\sigma_{k}^{+} + ( \alpha^{+}_{k} - \alpha^{+}_{ki})(\alpha^{-}_{k}\sigma_{k}^{+}  - \alpha^{-}_{ki}\sigma_{k}^{-})- \right. \\
& \left. - ( \alpha^{-}_{k} - \alpha^{-}_{ki} )(\alpha^{+}_{ki}\sigma_{k}^{+}  - \alpha^{+}_{k}\sigma_{k}^{-} ) \right] \sigma_{i}^{+},
\end{split}
\end{equation}
so that it is
\begin{equation}\label{e.sigma+_fin}
\begin{split}
& \dot{\sigma}_{i}^{+} = - \frac{\sigma_{i}^{+} }{2} + \frac{\alpha^{-}_{i}\alpha^{+}_{i}}{2} + \sum_{k \neq i}  \frac{1}{2} \left[ -\left( \alpha^{-}_{k} - \alpha^{-}_{ki} \right)^{2}n_{k}  - ( \alpha^{+}_{k} - \alpha^{+}_{ki} )^{2}\sigma_{k}^{-}\sigma_{k}^{+} + \right.\\
& \left. + ( \alpha^{+}_{k} - \alpha^{+}_{ki})(\alpha^{-}_{k}\sigma_{k}^{+}  - \alpha^{-}_{ki}\sigma_{k}^{-})- ( \alpha^{-}_{k} - \alpha^{-}_{ki} )(\alpha^{+}_{ki}\sigma_{k}^{+}  - \alpha^{+}_{k}\sigma_{k}^{-} ) \right] \sigma_{i}^{+}.
\end{split}
\end{equation}
In order to simplify last equation, we focus on fully-connected models with $E(\vec{z})=-\sum_{i,j}J_{ij}z_{i}z_{j}$, with the corresponding operator $\Delta E_{i}$ as defined in the main text. Requiring this energy to stay finite in the thermodynamic limit, we have $J_{ki} \sim 1 / N$. In the following, we perform a power series expansion with respect the parameter ($\beta J_{ki}$) neglecting terms of the order $O(1/N^{2})$. In particular, we consider the following expression
\begin{equation}
\begin{split}
& (\alpha^{+}_{k}-\alpha^{+}_{ki}) = \frac{e^{+\frac{\beta}{2}\sum_{j}J_{kj}\sigma_{j}^{z}}}{\sqrt{2\cosh(\beta \Delta E_{k})}} - \frac{e^{+\frac{\beta}{2}\sum_{j}J_{kj}\sigma_{j}^{z}} e^{-\beta J_{ki}} }{\sqrt{2\cosh( \beta\sum_{j \neq i}J_{kj}\sigma_{j}^{z} - \beta J_{ki})}} = \\
& =  \frac{e^{+\frac{\beta}{2}\sum_{j}J_{kj}\sigma_{j}^{z}}}{\sqrt{2\cosh(\beta \Delta E_{k})}} \left[ 1 - e^{-\beta J_{ki}} \left( \frac{ \cosh(\beta \Delta E_{k})}{ \cosh( \beta\sum_{j \neq i}J_{kj}\sigma_{j}^{z} - \beta J_{ki})} \right)^{\frac{1}{2}} \right] = \\
& = \frac{e^{+\frac{\beta}{2}\sum_{j}J_{kj}\sigma_{j}^{z}}}{\sqrt{2\cosh(\beta \Delta E_{k})}} \left[ 1 - e^{-\beta J_{ki}} \left( \frac{ \cosh(\beta \Delta E_{k})}{ \cosh( \beta\sum_{j}J_{kj}\sigma_{j}^{z} - \beta J_{ki}(1+\sigma_{i}^{z}))} \right)^{\frac{1}{2}} \right] \simeq \\
&  = \frac{e^{+\frac{\beta}{2}\sum_{j}J_{kj}\sigma_{j}^{z}}}{\sqrt{2\cosh(\beta \Delta E_{k})}} \left[ 1 - e^{-\beta J_{ki}} \left( \frac{1}{1-\tanh(\beta \sum_{j} J_{kj} \sigma_{j}^{z})\beta J_{ki}(1+\sigma_{i}^{z})}\right)^{\frac{1}{2}} \right] \simeq \\
& \frac{e^{+\frac{\beta}{2}\sum_{j}J_{kj}\sigma_{j}^{z}}}{\sqrt{2\cosh(\beta \Delta E_{k})}} \left[ 1 - (1- \beta J_{ki})(1 + \frac{1}{2} \tanh(\beta \sum_{jk} J_{kj} \sigma_{j}^{z}) )\beta J_{ki}(1+\sigma_{i}^{z}) \right] \simeq \\
&  \frac{e^{+\frac{\beta}{2}\sum_{j}J_{kj}\sigma_{j}^{z}}}{\sqrt{2\cosh(\beta \Delta E_{k})}} \left[ \beta J_{ki} -  \frac{1}{2} \tanh(\beta \sum_{jk} J_{kj} \sigma_{j}^{z}) \beta J_{ki}(1+\sigma_{i}^{z}) \right]  \simeq O(\beta J_{ki}).\\
\end{split}
 \end{equation}
Keeping terms up to the first order in $\beta J_{ki}$, then Eq.(\ref{e.sigma+_fin}) reads
\begin{equation}
\begin{split}
& \dot{\sigma}_{i}^{+} = - \frac{\sigma_{i}^{+} }{2} + \frac{\alpha^{-}_{i}\alpha^{+}_{i}}{2} +\sum_{k\neq i}(\beta J_{k i})\alpha^{-}_{k}\alpha^{+}_{k}(i\sigma_{k}^{y})\sigma_{i}^{+} = \\
& - \frac{\sigma_{i}^{+} }{2} + \frac{1}{4}\mathrm{sech}( \beta \Delta E_{i}) + i \frac{\beta}{2}\sum_{k \neq i }\mathrm{sech}(\beta \Delta E_{k})\sigma_{k}^{y}\sigma_{i}^{+}, \\
\end{split}
\end{equation}
and the dynamical equations for $ \sigma_{i}^{x,y} $ are

\begin{equation}\label{e.dynamical}
\begin{split}
 \dot{\sigma}_{i}^{x} = & - \frac{\sigma_{i}^{x} }{2} + \frac{1}{2}\mathrm{sech}( \beta \Delta E_{i})-\frac{\beta}{2}\sum_{k\neq i}J_{k i}\mathrm{sech}(\beta \Delta E_{k})\sigma_{k}^{y}\sigma_{i}^{y},\\
\dot{\sigma}_{i}^{y} = & - \frac{\sigma_{i}^{y} }{2} + \frac{\beta}{2}\sum_{k\neq i}J_{k i}\mathrm{sech}(\beta \Delta E_{k})\sigma_{k}^{y}\sigma_{i}^{x},\\
\end{split}
\end{equation}

\subsection{Unitary transformation of jump operators}

As we stress in the main text, the stationary state of the dynamics given by Eq.(\ref{B-eq.Lindblad}) does not change under any local set of unitary transformation $u_{i}$, ($u_{i}^{\dagger}u_{i} = \mathbf{1}$), that modifies the set $L_{i}$ as follows

\begin{equation}
L_{i}^{\prime} \equiv u_{i}L_{i} = \alpha^{-}_{i} u_{i}n_{i} - \alpha^{+}_{i} u_{i}\sigma_{i}^{+},
\end{equation}
We choose to generalize the jump operators with the following equal spin rotation on all sites
\begin{equation}
u_{i} = \begin{pmatrix} \cos\theta & e^{i \phi}\sin\theta \\
 -e^{-i\phi}\sin\theta & \cos\theta \end{pmatrix}.
\end{equation}
In order to obtain the equations of motion for the operator $\sigma_{i}^{z}$, it is useful to evaluate the matrix representation of the operator $ L_{i}^{\dagger}L_{i}$, which is
\begin{equation}
L_{i}^{\dagger}L_{i} = \begin{pmatrix}
\alpha^{-}_{i} & 0 \\
-\alpha^{+}_{i} & 0
\end{pmatrix} 
\begin{pmatrix}
\alpha^{-}_{i} & -\alpha^{+}_{i}\\
0 & 0
\end{pmatrix}
= \begin{pmatrix}
(\alpha^{-}_{i})^{2} & -\alpha^{-}_{i}\alpha^{+}_{i} \\
-\alpha^{-}_{i}\alpha^{+}_{i} & (\alpha^{+}_{i})^{2}
\end{pmatrix}.
\end{equation}
Le us start by considering the dynamical equations for $\sigma_{i}^{z}$, which read
\begin{equation}
\dot{\sigma}_{i}^{z} = L_{i}^{\prime \dagger}\sigma_{i}^{z}L_{i}^{\prime} - \frac{1}{2}\lbrace L_{i}^{\prime \dagger}L_{i}^{\prime}, \sigma_{i}^{z} \rbrace = L_{i}^{\prime \dagger}\sigma_{i}^{z}L_{i}^{\prime} - \frac{1}{2}\lbrace L_{i}^{ \dagger}L_{i}, \sigma_{i}^{z} \rbrace,
\end{equation}
where the first term on the right-hand side reads
\begin{equation}
 L_{i}^{\prime \dagger}\sigma_{i}^{z}L_{i}^{\prime} = L_{i}^{\dagger}u_{i}^{\dagger}\sigma_{i}^{z}u_{i}L_{i} = [2\cos(\theta)^{2}-1]L_{i}^{\dagger}L_{i} = f_{z}L_{i}^{\dagger}L_{i},
\end{equation}
with the definition $f_{z}(\theta) \equiv  \cos(2\theta)$. Thus, it is
\begin{equation}
\dot{\sigma}_{i}^{z}  = -f_{z}\left[ \alpha^{-}_{i}\alpha^{+}_{i}\sigma_{i}^{x} - \frac{1}{2}((\alpha^{-}_{i})^{2} - (\alpha^{+}_{i})^{2})\sigma_{i}^{z} - \frac{1}{2} \right] - \frac{1}{2}((\alpha^{-}_{k})^{2} - (\alpha^{+}_{k})^{2}) - \frac{1}{2}\sigma_{k}^{z}.
\end{equation}
The dynamical equation for $\sigma_{i}^{+}$ in the thermodynamic limit is
\begin{equation}
\dot{\sigma}_{i}^{+} = L_{i}^{\prime \dagger}\sigma_{i}^{+}L_{i}^{\prime} - \frac{1}{2}\lbrace L_{i}^{ \dagger}L_{i}, \sigma_{i}^{+} \rbrace + \sum_{k \neq i}L_{k }^{\prime \dagger}\sigma_{i}^{+}L^{ \prime}_{k}-\frac{1}{2}\lbrace L_{k}^{\dagger}L_{k}, \sigma_{i}^{+} \rbrace,
\end{equation}
where
\begin{equation}
\begin{split}
& L_{i}^{\prime \dagger}\sigma_{i}^{+}L_{i}^{\prime} = L_{i}^{\dagger}u_{i}^{\dagger}\sigma_{i}^{+}u_{i}L_{i} = -\cos(\theta)\sin(\theta)e^{-i\phi} L_{i}^{\dagger}L_{i}, \\
& L_{k}^{\prime \dagger}\sigma_{i}^{+}L_{k}^{\prime} = L_{k}^{\dagger}u_{k}^{\dagger}\sigma_{i}^{+}u_{k}L_{k} = L_{k}^{ \dagger}\sigma_{i}^{+}L_{k}.
\end{split} 
\end{equation}
Hence we get
\begin{equation}
\begin{split}
& \dot{\sigma}_{i}^{+} = \cos(\theta)\sin(\theta)e^{-i\phi} \left[ \alpha^{-}_{i}\alpha^{+}_{i}\sigma_{i}^{x} -\frac{((\alpha^{-}_{i})^{2} + (\alpha^{+}_{i})^{2})}{2} -\frac{((\alpha^{-}_{i})^{2} - (\alpha^{+}_{i})^{2})}{2}\sigma_{i}^{z} \right] - \\
& - \frac{((\alpha^{-}_{i})^{2} + (\alpha^{+}_{i})^{2})}{2}\sigma_{i}^{+} + \frac{\alpha^{-}_{i}\alpha^{+}_{i}}{2}+ i \frac{\beta}{2}\sum_{k \neq i }\mathrm{sech}(\beta \Delta E_{k})\sigma_{k}^{y}\sigma_{i}^{+}. \\
\end{split}
\end{equation}
The equation of motion of the operators $\sigma_{i}^{x,y}$ are
\begin{equation}
\begin{split}
& \dot{\sigma}_{i}^{x} = - f_{x} \left[ \alpha^{-}_{i}\alpha^{+}_{i}\sigma_{i}^{x} - \frac{((\alpha^{-}_{i})^{2} + (\alpha^{+}_{i})^{2})}{2} -\frac{((\alpha^{-}_{i})^{2} - (\alpha^{+}_{i})^{2})}{2}\sigma_{i}^{z} \right] \\
& \quad - \frac{((\alpha^{-}_{i})^{2} + (\alpha^{+}_{i})^{2})}{2}\sigma_{i}^{x} + \alpha^{-}_{i}\alpha^{+}_{i} - \frac{\beta}{2}\sum_{k \neq i }\mathrm{sech}(\beta \Delta E_{k})\sigma_{k}^{y}\sigma_{i}^{y}, \\
& \dot{\sigma}_{i}^{y} = - f_{y} \left[ \alpha^{-}_{i}\alpha^{+}_{i}\sigma_{i}^{x} - \frac{((\alpha^{-}_{i})^{2} + (\alpha^{+}_{i})^{2})}{2} -\frac{((\alpha^{-}_{i})^{2} - (\alpha^{+}_{i})^{2})}{2}\sigma_{i}^{z} \right] \\
& \quad - \frac{((\alpha^{-}_{i})^{2} + (\alpha^{+}_{i})^{2})}{2}\sigma_{i}^{y} + \frac{\beta}{2}\sum_{k \neq i }\mathrm{sech}(\beta \Delta E_{k})\sigma_{k}^{x}\sigma_{i}^{y}, \\
\end{split}
\end{equation}
where $f_{x}(\theta, \phi) \equiv  -\sin(2\theta)\cos(\phi)$ and $f_{y}(\theta, \phi) \equiv \sin(2\theta)\sin(\phi) $, and the dynamical equations are

\begin{equation}\label{sm.spin_gen}
\begin{split}
 \dot{\sigma}_{i}^{x} =&-A_{i}^{x}(\beta) -\frac{\beta}{2}\sum_{k \neq i } J_{k i} \mathrm{sech}(\beta \Delta E_{k}) \sigma_{k}^{y} \sigma_{i}^{y}+ \frac{ \mathrm{sech}(\beta \Delta E_{i})}{2}, \\
 \dot{\sigma}_{i}^{y} =&-A_{i}^{y}(\beta) +\frac{\beta}{2}\sum_{k \neq i } J_{k i} \mathrm{sech}(\beta \Delta E_{k}) \sigma_{k}^{y}\sigma_{i}^{x},\\
 \dot{\sigma}_{i}^{z} =& -A_{i}^{z}(\beta)+ \frac{1}{2}\tanh(\beta \Delta E_{i}),\\
\end{split} 
\end{equation}
where we have defined $A_{i}^{\alpha}(\beta)=\lbrace f_{\alpha}\left[\mathrm{sech}(\beta \Delta E_{i}) \sigma_{i}^{x}+\tanh(\beta \Delta E_{i})\sigma_{i}^{z} -1\right]+\sigma_{i}^{\alpha}\rbrace/2$, $\alpha=x,y,z$, which parametrically depends on the angles $(\theta,\phi)$, by means of the functions $f_{x,y,z}(\theta, \phi)$. We observe that if one takes $f_{x}=0$, i.e. $\theta_{i}=\frac{\pi}{4} \; \forall i$, the equation for $\sigma_{z}^{i}$ decouples from the other operators and reads $\dot{\sigma}_{i}^{z} = \frac{1}{2}[ - \sigma_{i}^{z} + \tanh(\beta \Delta E_{i})]$. This allows us to make the comparison between the quantum dynamics and the classical one.

\section{Fully connected Ising model}
\label{s.sec3.2}

The case with uniform couplings $J_{ij}=1/N$, $\forall i,j$ allows us to compare the quantum dynamics with the one of the fully connected Ising model whose classical energy function is $E=-1/N\sum_{i,j}z_{i}z_{j}$. In the thermodynamic limit, the $3N$ equations of motions (\ref{sm.spin_gen}) are reduced to the closed set of $3$ equations (\ref{B-e.ferromagnetic}) for the collective operators $m^{\alpha}=\braket{\hat{s}^{\alpha}}=1/N \sum_{i=1}^{N}\braket{\sigma_{i}^{\alpha}}$, $\alpha=x,y,z$, as reported in the main text.

\subsection{Stationary condition}

For the reader's convenience we repeat here the set of equations \eqref{B-e.ferromagnetic} from the main text:
\begin{equation}
\begin{split}
 \dot{m}^{x}=&-A_{x} + \frac{1}{2}\mathrm{sech}(\beta m^{z}) -\frac{\beta}{2} \mathrm{sech}(\beta m^{z}) (m^{y})^{2},\\
 \dot{m}^{y}=&-A_{y} + \frac{\beta}{2} \mathrm{sech}(\beta m^{z}) m^{y} m^{x},\\
 \dot{m}^{z}=&-A_{z} + \frac{1}{2}\tanh(\beta m^{z}), \\
\end{split}
\label{sm.ferromagnetic}
\end{equation}
where $A_{\alpha} = \lbrace f_{\alpha}\left[ m^{x}\mathrm{sech}(\beta m^{z}) + m^{z}\tanh(\beta m^{z}) -1 \right]+m^{\alpha}\rbrace/2 $, $\alpha=x,y,z$ and
\be
	f_x = -\sina{2\theta} \cosa{\phi} \comma f_y = \sina{2\theta} \sina{\phi} \comma f_z = \cosa{2\theta}.
\ee
We first check the physical consistency of the dynamics described by these equations showing that the portion of phase space outside the unit sphere
\be
	R^2 \equiv (m^x)^2 + (m^y)^2 + (m^z)^2 = 1
\ee
is transient, i.e., the dynamics is eventually confined to $R \leq 1$ for any initial condition. To do this, we consider the derivative of the ``radius'' itself, i.e.
\be
	 \partial_t R^2 = 2 (m^x \dot{m}^x + m^y \dot{m}^y + m^z \dot{m}^z). 
	 \label{sm.dtR}
\ee
For future convenience, we separate the two addends inside the curly brackets of $A^{x,y,z}$ and introduce the shorthand
\be
	F = 1 - m^{x}\mathrm{sech}(\beta m^{z}) - m^{z}\tanh(\beta m^{z}).
\ee
Hence, by substitution of \eqref{sm.ferromagnetic} into \eqref{sm.dtR} we obtain
\be
	\partial_t R^2 = F (f_x m^x + f_y m^y + f_z m^z) - R^2 + \underbrace{m^z \tanh(\beta m^z) + m^x \mathrm{sech}(\beta m^z)}_{= 1-F},
\ee
i.e.,
\be
	\partial_t R^2 = F (f_x m^x + f_y m^y + f_z m^z - 1) + (1- R^2).
	\label{sm.dtR2}
\ee
We now focus on $F$ itself to obtain a bound on the values it can take; more precisely, we consider
\be
	\abs{1-F} = \abs{ m^{x}\mathrm{sech}(\beta m^{z}) + m^{z}\tanh(\beta m^{z})} = \abs{\vec{v} \cdot \vec{R}},
\ee
where we introduced two effective vectors
\be
	\vec{R} = (m^x, m^y, m^z)^T \mand \vec{v} = (\mathrm{sech}(\beta m^{z}), 0 ,\tanh(\beta m^{z}))^T.
\ee
By Cauchy-Schwarz inequality we then have
\be
	\abs{1-F} \leq \underbrace{\abs{\vec{v}}}_{=1} \abs{\vec{R}} = R.
\ee
This then implies 
\be
	1-R \leq F \leq 1 + R.
	\label{sm.bound1}
\ee
In an analogous fashion, we see that $\abs{f_x m^x + f_y m^y + f_z m^z} = \abs{\vec{f} \cdot \vec{R}}$ with $\vec{f} = (f_x , f_y ,f_z )^T$ another unit vector. Hence,
\be
	\abs{f_x m^x + f_y m^y + f_z m^z} \leq R 
\ee
and
\be
	-R - 1 \leq f_x m^x + f_y m^y + f_z m^z - 1 \leq R - 1.
	\label{sm.bound2}
\ee

We now assume $R > 1$ (placing ourselves outside the unit sphere in phase space). In Eq.~\eqref{sm.dtR2} we can now use the bounds \eqref{sm.bound1} and \eqref{sm.bound2} to determine the sign of the derivative. In fact, for $R > 1$ the upper (lower) bounds are both positive (negative), so that the first addend in \eqref{sm.dtR2} is always smaller than any of the two products $(1-R)(-R - 1)$ or $(1+R)(R-1)$. Since they both yield the same result, we can summarise the result by
\be
	\partial_t R^2 \leq (1+R)(R-1) + (1- R^2) = 0,
\ee
which shows that, for any $R>1$, the evolution is either ingoing or tangential to the sphere of radius $R$. The cases of tangential evolution are either stationary points or the trajectory will then move to inward-going points. In the absence of stationary points for $R>1$ the portion of phase space with $R>1$ must be transient. We thereby move now onto proving there are no such steady states. First, we note that $\partial_t R^2 = 0$ only if we can replace our inequalities above with equalities. This implies
\be
	\vec{v} = \vec{f} = \pm \frac{\vec{R}}{R},
\ee
with $\vec{v}$ and $\vec{f}$ defined above. In particular, this yields
\be
	m^y = f_y = 0 \comma \mathrm{sech}(\beta m^{z}) = f_x  \mand \tanh(\beta m^z) = f_z .
\ee

\subsection{Stability analysis}

We analyse the time-scales that characterize the long-time approach of $m^{\alpha}$ to the "ferromagnetic" stationary solution that appears for $\beta > 1$, comparing the quantum case (generic $\theta$ and $\phi$) with the classical one ($\theta=\pi/4$). As we are considering the dynamics near to the stationary solution, we perform a first order expansion of Eqs.(\ref{B-e.ferromagnetic}) around the fixed points, say $\bar{m^{\alpha}}$, so that $m^{\alpha} \sim \bar{m}^{\alpha} + \delta m^{\alpha}$. It is worth noting that the stationary solution itself is independent on the unitary transformation that has been applied on the jump operators, i.e. on the angles $\theta$ and $\phi$, hence we can get the stationary solution $\bar{m}^{\alpha}$ fixing the values of $\theta$ and $\phi$. For the sake of simplicity, we set $\theta=\pi/4$, $\phi=0$ getting  $\bar{m}^{x}= \mathrm{sech}(\beta \bar{m}^{z})$, $\bar{m}^{y}=0$ and $\bar{m}^{z}= \tanh(\beta \bar{m}^{z})$. 

The equations for variables $\delta m^{\alpha}$, $\alpha = x,y,z$ can be written as follows,

\begin{equation}
\begin{pmatrix}
\delta \dot{m}^{x} \\
\delta \dot{m}^{y} \\
\delta \dot{m}^{z} \\
\end{pmatrix} = S 
\begin{pmatrix}
\delta m^{x} \\
\delta m^{y} \\
\delta m^{z} \\
\end{pmatrix}.
\end{equation} 
from which we see that the long-time decay towards the stationary state is exponential with the rates given by the eigenvalues of the matrix $S$, whose elements $S_{ij}$ are
\begin{equation}\label{e.linear}
\begin{split}
S_{11}= & -\frac{1}{2}(f_{x}\mathrm{sech}(\beta  \bar{m}^{z})+1),\\
S_{12}= & 0 \\
S_{13} = & \frac{1}{4}\mathrm{sech}^{2}(\beta \bar{m}^{z})\left[f_{x}(2\beta  \bar{m}^{x}\sinh(\beta  \bar{m}^{z})-2\beta  \bar{m}^{z}-\sinh(2\beta  \bar{m}^{z}))-2\beta \sinh(\beta  \bar{m}^{z})\right],\\ 
S_{21}= & -\frac{f_{y}}{2} \mathrm{sech}(\beta \bar{m}^{z}),\\
S_{22}= &\frac{1}{2}[-1+\beta \mathrm{sech}(\beta \bar{m}^{z})\bar{m}^{x}],\\
S_{23}= & \frac{f_{y}}{4}\mathrm{sech}^{2}(\beta \bar{m}^{z})(2\beta  \bar{m}^{x}\sinh(\beta  \bar{m}^{z})-2\beta  \bar{m}^{z}-\sinh(2\beta  \bar{m}^{z}) )\\
S_{31} = &-\frac{f_{z}}{2}\mathrm{sech}(\beta  \bar{m}^{z}), \\
S_{32} = & 0, \\
S_{33}= & \frac{\mathrm{sech}^{2}(\beta \bar{m}^{z})}{4} \left[f_{z}(2\beta  \bar{m}^{x}\sinh(\beta  \bar{m}^{z})-2\beta  \bar{m}^{z}-\sinh(2\beta  \bar{m}^{z})) -2\beta \right] -\frac{1}{2}. \\ 
\end{split}
\end{equation}
We can immediately see that when substituting the stationary values of $\bar{m}^{x}$ in $S_{22}$, which is an eigenvalue of $S$, this is equal to the classical one obtained fixing $\theta=\pi/4$. This implies that, near enough to the stationary solutions, the quantum dynamics is dominated by the same time-scale or by a longer time-scale than the classical one. Therefore, the faster approach towards the stationary state of the quantum case over the classical one, which occurs for some value of $\theta$ and $\phi$ as reported in the main text, is only due by the early stages of the dynamics.


\section{Dynamical equations for the p memory case}

We tackle now the problem of getting the Eqs.(\ref{B-e.hopfieldgen}). The dynamical equations for the overlap between the spin configuration and the $\mu$-th pattern, $\hat{s}_{\mu}^{\alpha} \equiv \frac{1}{N} \sum_{i=1}^{N} \xi_{i}^{(\mu)} \sigma_{i}^{\gamma},  \alpha = x,y,z$, read
\begin{equation}\label{e.overlap_uni}
\begin{split}
& \dot{\hat{s}}_{\mu}^{x} = -\frac{f}{2N}\left[ \sum_{i}\xi_{i}^{(\mu)} \mathrm{sech}( \beta \vec{\xi}_{i} \cdot \vec{s}_{z} )\sigma_{i}^{x} +\sum_{i}\xi_{i}^{(\mu)}  \tanh( \beta \vec{\xi}_{i} \cdot \vec{s}_{z})\sigma_{i}^{z} -\sum_{i}\xi_{i}^{(\mu)} \right] \\
&-\frac{1}{2}\hat{s}_{\mu}^{x} + \frac{1}{2N} \sum_{i} \xi_{i}^{(\mu)} \mathrm{sech}( \beta \vec{\xi}_{i} \cdot \vec{s}_{z}) - \frac{\beta}{2N^{2}} \sum_{i}\sum_{k\neq i}\sum_{\tilde{\mu}}\xi_{i}^{\mu} \xi_{i}^{\tilde{\mu}} \xi_{k}^{\tilde{\mu}} \mathrm{sech}(\beta \vec{\xi}_{k} \cdot \vec{s}_{z}) \sigma_{k}^{y}\sigma_{i}^{y}, \\
& \dot{\hat{s}}_{\mu}^{y} = -\frac{h}{2N}\left[ \sum_{i}\xi_{i}^{(\mu)} \mathrm{sech}( \beta \vec{\xi}_{i} \cdot \vec{s}_{z} )\sigma_{i}^{x} +\sum_{i}\xi_{i}^{(\mu)}  \tanh( \beta \vec{\xi}_{i} \cdot \vec{s}_{z})\sigma_{i}^{z} -\sum_{i}\xi_{i}^{(\mu)} \right] \\
&-\frac{1}{2}\hat{s}_{\mu}^{y} + \frac{\beta}{2N^{2}} \sum_{i}\sum_{k\neq i}\sum_{\tilde{\mu}}\xi_{i}^{\mu} \xi_{i}^{\tilde{\mu}} \xi_{k}^{\tilde{\mu}} \mathrm{sech}(\beta \vec{\xi}_{k} \cdot \vec{s}_{z}) \sigma_{k}^{y}\sigma_{i}^{x}, \\
& \dot{\hat{s}}_{\mu}^{z} = -\frac{g}{2N}\left[ \sum_{i}\xi_{i}^{(\mu)} \mathrm{sech}( \beta \vec{\xi}_{i} \cdot \vec{s}_{z} )\sigma_{i}^{x} +\sum_{i}\xi_{i}^{(\mu)}  \tanh( \beta \vec{\xi}_{i} \cdot \vec{s}_{z})\sigma_{i}^{z} -\sum_{i}\xi_{i}^{(\mu)} \right] \\
& \qquad \quad + \frac{1}{2N}\sum_{i}\xi_{i}^{(\mu)} \tanh( \beta \vec{\xi}_{i} \cdot \vec{s}_{z})   - \frac{1}{2}\hat{s}_{\mu}^{z},
\end{split}
\end{equation}
which do not easily close on the overlaps variables. However, we point out that (i) the stationary solution is independent on the unitary transformation, therefore it is the same of the classical case that one obtains by considering $\theta=\pi/4$; (ii) the stationary dark state reproduces in its diagonal elements the probabilities distribution of a classical Hopfiled NN as defined at the beginning; (iii) given any operator $O(\sigma^{z})$ that is diagonal on the $\sigma_{i}^{z}$ basis, the expectation value on the stationary state is the classical average value, being the correspondent weights the thermal ones, i.e. $\braket{O(\sigma_{z})} = \mathrm{Tr}(\rho_{SS}O(\sigma_{z}))=\frac{1}{Z(\beta)} \sum_{\vec{\sigma}}e^{-\beta E(\vec{\sigma})}\braket{\vec{\sigma}|O(\sigma^{z})|\vec{\sigma}}$. 

The previous considerations lead us to argue that the dynamics of the expectation value of spin operators will be mostly determined by the properties of the overlap and therefore we employ the following approximation, 
\begin{equation}
\braket{\sigma_{i}^{\alpha}} \sim \xi_{i}^{\mu}m_{\mu}^{\alpha}, \; \alpha=x,y,z,
\end{equation}
where $m_{\mu}^{\alpha}=\braket{s_{\mu}^{\alpha}}$ represents the expectation value. Performing a mean field approximation, i.e. neglecting the correlations among the overlaps, and employing a self average hypothesis we get
\begin{equation}\label{sm.hopfield}
\begin{split}
& \dot{m}_{\mu}^{x,y}= -\bar{A}_{\mu}^{x,y}(\beta), \\
& \dot{m}_{\mu}^{z}= -\bar{A}_{\mu}^{z}(\beta) + \frac{1}{2}  \overline{\xi^{\mu}\tanh{\beta \vec{\xi} \cdot \vec{m}^{z}}},\\
\end{split}
\end{equation}
where $ \bar{A}_{\mu}^{\alpha}= \left[f_{\alpha} m_{\mu}^{x} \overline{ \mathrm{sech}(\beta \vec{\xi} \cdot \vec{m}^{z})}+m_{\mu}^{\alpha}\right]/2$, $\alpha = x,y,z$, $\vec{\xi}=(\xi^{1},...,\xi^{p})^{T}$, $\vec{m}^{\alpha}=(m^{\alpha}_{1},...,m^{\alpha}_{p})^{T}$, and we replaced $\frac{1}{N} \sum_{i}f(\vec{\xi}_{i})\rightarrow \overline{f(\vec{\xi})}$, $\overline{(\cdot)}$ being the average over the disorder distribution.

As a first consistency check for the approximations employed, we have verified that the stationary state is invariant under change of the angles $(\theta, \phi)$ and that its qualitative structure is consistent with the classical one, i.e., only a paramagnetic stable solution appears at $\beta < 1$, whereas several
retrieval solutions emerge at $\beta > 1$. 

Second, we checked the results of these equations for small $\beta$ with a high temperature expansion of Eqs.(\ref{sm.hopfield}), obtaining agreement both in the early and late stages of the dynamics. In order to perform the high temperature expansion we notice that, in the limit $\beta \rightarrow 0$, the jump operator defined in Eq.(\ref{B-eq.jumpop}) for the site $i$ is independent on the other spins and leads the system in a state where the magnetization $M^{x}=\frac{1}{N}\sum_{i}\sigma_{i}^{x}$ takes its maximum value. In this regime, closing the equations both on the overlaps and the magnetizations seems a consistent choice. If this is the case, we can expand the equations in $\beta$, employing the following approximations
\begin{equation}
\frac{1}{N}\sum_{i}\xi_{i}^{\nu}\xi_{i}^{\gamma}\sigma_{i}^{\alpha} \approx  \delta_{\nu \gamma} M^{\alpha} \qquad \frac{1}{N}\sum_{i}\xi_{i}^{\nu}\xi_{i}^{\gamma}\xi_{i}^{\rho}\sigma_{i}^{\alpha} \approx m^{\alpha}_{\nu}\delta_{\gamma \rho} + \mathrm{perm}.
\end{equation}
Expanding up to the third order in $\beta$ we get
\begin{equation}
\begin{split}
 \dot{M}^{x}= &-\frac{f_{x}}{2}\left\lbrace M^{x} -1 +\beta |\vec{m}^{z}|^{2} - \frac{\beta^{2}}{2}M^{x}|\vec{m}^{z}|^{2}-\frac{\beta^{3}}{3}\left[3|\vec{m}^{z}|^{4}-2\sum_{\nu}\left(m_{\nu}^{z} \right)^{4} \right] \right\rbrace - \frac{M^{x}}{2} + \\
& +\frac{1}{2} -\frac{\beta}{2} |\vec{m}^{y}|^{2}  - \frac{\beta^{2}}{4}|\vec{m}^{z}|^{2} + \frac{\beta^{3}}{4}\left[2 (\vec{m}^{z}\cdot \vec{m}^{y})^{2} + |\vec{m}^{z}|^{2} |\vec{m}^{y}|^{2} - 2\sum_{\nu}(m_{\nu}^{z}m_{\nu}^{y})^{2}  \right], \\
\dot{M}^{y}= & -\frac{f_{y}}{2}\left\lbrace M^{x} -1 +\beta |\vec{m}^{z}|^{2} - \frac{\beta^{2}}{2}M^{x}|\vec{m}^{z}|^{2}-\frac{\beta^{3}}{3}\left[3|\vec{m}^{z}|^{4}-2\sum_{\nu}\left(m_{\nu}^{z} \right)^{4} \right] \right\rbrace - \frac{M^{y}}{2} + \\
& + \frac{\beta}{2}(\vec{m}^{y}\cdot \vec{m}^{x})^{2} - \frac{\beta^{3}}{4} \left[ 2 (\vec{m}^{z}\cdot \vec{m}^{y})(\vec{m}^{z}\cdot \vec{m}^{x})+  |\vec{m}^{z}|^{2} (\vec{m}^{y}\cdot \vec{m}^{x}) -2 \sum_{\nu}(m_{\nu}^{z})^{2}m_{\nu}^{x}m_{\nu}^{y} \right], \\
\dot{M}^{z}= & -\frac{f_{z}}{2}\left\lbrace M^{x} -1 +\beta |\vec{m}^{z}|^{2} - \frac{\beta^{2}}{2}M^{x}|\vec{m}^{z}|^{2}-\frac{\beta^{3}}{3}\left[3|\vec{m}^{z}|^{4}-2\sum_{\nu}\left(m_{\nu}^{z} \right)^{4} \right] \right\rbrace - \frac{M^{z}}{2}; \\
\end{split}
\end{equation}
\begin{equation}\label{e.expansion_beta}
\begin{split}
 \dot{m}_{\mu}^{x}= & -\frac{f_{x}}{2} \left\lbrace  m_{\mu}^{x} + \beta M^{z} m_{\mu}^{z} - \frac{\beta^{2}}{2} \left[ m_{\mu}^{x}|\vec{m}^{z}|^{2} + 2m_{\mu}^{z}(\vec{m}^{z} \cdot \vec{m}^{x})-2m_{\mu}^{x}(m_{\mu}^{z})^{2} \right] + \right. \\
& \left. -\frac{\beta^{3}}{3}M^{z}\left[3m_{\mu}^{z}|\vec{m^{z}}|^{2} - 2(m_{\mu}^{z})^{3} \right] \right\rbrace -\frac{m_{\mu}^{x}}{2} -\frac{\beta}{2}M^{y}m_{\mu}^{y} + \\
& + \frac{\beta^{3}}{4}M^{y}\left[2m_{\mu}^{z}(\vec{m}^{z} \cdot \vec{m}_{\mu}^{y}) + m_{\mu}^{y}|m^{z}|^{2}-2m_{\mu}^{y}(m_{\mu}^{z})^{2} \right],\\
 \dot{m}_{\mu}^{y}= & -\frac{f_{y}}{2} \left\lbrace  m_{\mu}^{x} + \beta M^{z} m_{\mu}^{z} - \frac{\beta^{2}}{2} \left[ m_{\mu}^{x}|\vec{m}^{z}|^{2} + 2m_{\mu}^{z}(\vec{m}^{z} \cdot \vec{m}^{x})-2m_{\mu}^{x}(m_{\mu}^{z})^{2} \right] + \right. \\
& \left. -\frac{\beta^{3}}{3}M^{z}\left[3m_{\mu}^{z}|\vec{m^{z}}|^{2} - 2(m_{\mu}^{z})^{3} \right] \right\rbrace -\frac{m_{\mu}^{y}}{2} +\frac{\beta}{2}M^{x}m_{\mu}^{y} + \\
& + \frac{\beta^{3}}{4}M^{x}\left[2m_{\mu}^{z}(\vec{m}^{z} \cdot \vec{m}_{\mu}^{y}) + m_{\mu}^{y}|m^{z}|^{2}-2m_{\mu}^{y}(m_{\mu}^{z})^{2} \right],\\
 \dot{m}_{\mu}^{z}= & -\frac{f_{z}}{2} \left\lbrace  m_{\mu}^{x} + \beta M^{z} m_{\mu}^{z} - \frac{\beta^{2}}{2} \left[ m_{\mu}^{x}|\vec{m}^{z}|^{2} + 2m_{\mu}^{z}(\vec{m}^{z} \cdot \vec{m}^{x})-2m_{\mu}^{x}(m_{\mu}^{z})^{2} \right] + \right. \\
& \left. -\frac{\beta^{3}}{3}M^{z}\left[3m_{\mu}^{z}|\vec{m^{z}}|^{2} - 2(m_{\mu}^{z})^{3} \right] \right\rbrace -\frac{m_{\mu}^{z}}{2} -\frac{\beta}{2}m_{\mu}^{z} -\frac{\beta^{3}}{6}\left[ 3m_{\mu}^{z}|m^{z}|^{2}-2(m_{\mu}^{z})^{3} \right]. \\
\end{split}
\end{equation}
Comparing Eqs.(\ref{sm.hopfield}) and (\ref{e.expansion_beta}) we find that the former converge to the latter as $\beta$ decreases, not only near to the stationary solution but also in the transient part of the evolution. Such result make us argue that, although Eqs.(\ref{sm.hopfield}) are only a possible approximation of Eqs.(\ref{e.overlap_uni}), we can as a first step analyse them, near to the stationary solutions as well as in the transient regime of the dynamics. 

\subsection{Stability analysis}

Finally, we perform the stability analysis of the Eqs.(\ref{sm.hopfield}), expanding them around the stationary solutions of the overlaps for $\beta>1$ up to the first order. For the sake of simplicity, we fix $\theta=\pi/4$, $\phi=\pi/2$ in order to get the stationary solutions which read $\bar{m}_{\nu}^{x,y}=0$, $\bar{m}_{\nu}^{z}=\overline{\xi^{\nu}\tanh(\beta \vec{\xi} \cdot \vec{\bar{m}}^{z})}$. The expressions of the overlaps are thus taken to be 
\begin{equation}\tag{S32}
\begin{split}
& m_{\nu}^{x,y} \approx \delta m_{\nu}^{x,y} \quad \forall \nu, \\
& m_{\mu}^{z} \approx \bar{m}_{\mu}^{z} + \delta m_{\mu}^{z}, \quad m_{\nu}^{z} \approx \delta m_{\nu}^{z} \: \forall \nu \neq \mu,\\
\end{split}
\end{equation}
and we get
\begin{equation}\tag{S33}
\begin{pmatrix}
\delta \dot{m}_{\mu}^{x} \\
\delta \dot{m}_{\mu}^{y} \\
\delta \dot{m}_{\mu}^{z} \\
\end{pmatrix} =
\begin{pmatrix}
-\frac{1}{2}\left[ f_{x}\mathrm{sech}(\beta \bar{m}_{\mu}^{z}) +1 \right]  && 0 && 0 \\
-\frac{f_{y}}{2}\mathrm{sech}(\beta \bar{m}_{\mu}^{z}) && -\frac{1}{2} && 0 \\
-\frac{f_{z}}{2}\mathrm{sech}(\beta \bar{m}_{\mu}^{z}) && 0 && \frac{1}{2}\left[\frac{\beta}{\cosh^{2}(\beta \bar{m}_{\mu}^{z})}-1\right]\\
\end{pmatrix} 
\begin{pmatrix}
\delta m_{\mu}^{x} \\
\delta m_{\mu}^{y} \\
\delta m_{\mu}^{z} \\
\end{pmatrix}.
\end{equation}
As in the previous case, one of the eigenvalue of the matrix is the same of the classical case obtained for $\theta=\pi/4$. Therefore, the quantum time-scale that dominates the dynamics near to the stationary solutions is the same or longer than the classical one: the gain that we get from the quantum dynamics over the classical one arises only from the transient time-scales.

\subsection{Symmetries within the $\theta$-$\phi$ plane}

We discuss here the three symmetries we exploit which allow us to focus on a portion (one eighth) of the $\theta$-$\phi$ plane in Fig.~\ref{B-fig3} of the main text. The first one is a symmetry of the dynamics, which is invariant under the involutive transformation $(\theta, \phi) \to (\pi - \theta, \phi + \pi)$ (remembering that $\theta$ is defined modulo $\pi$ and $\phi$ modulo $2 \pi$). For any given choice of the initial conditions, the dynamics at $(\theta, \phi)$ exactly matches the one at $(\pi - \theta, \phi + \pi)$ and the timescales involved are therefore also the same. 

The other symmetries of the plot are not symmetries of the dynamics, and are only recovered upon averaging over the initial conditions. The first one involves the substitution $\phi \to \pi -\phi$ (which changes the sign of $f_y$) and $\vec{m}^y \to -\vec{m}^y$, which leaves the equations invariant. Therefore, any trajectory $\vec{m}^{x,y,z} (t)$ obtained at $(\theta, \phi)$ is equivalent to one obtained at $(\theta, \pi - \phi)$ where the $y$ component is simply inverted. Clearly, this does not affect the timescales of the dynamics. Since we are sampling over a symmetric distribution of initial conditions on $\vec{m}^y$, averages must be symmetric under $\phi \to \pi - \phi$.

Finally, we consider the transformation $\theta \to \pi/2 - \theta$ and $\vec{m}^z \to - \vec{m}^z$, which also leaves the equations invariant. For ``quantum'' trajectories the argument follows the same steps employed above: since the initial condition is generated symmetrically in $m^z$, averages should be symmetric under $\theta \to \pi/2 - \theta$. For the classical trajectories, since the initial conditions are biased by a fixed amount $Z_0$ this does not hold. However, the classical case ($\theta = \pi/4$) is special since $\theta$ gets simply mapped onto itself (and in fact $\vec{m}^z \to - \vec{m}^z$ represents by itself a symmetry of the equations). Hence, the average dynamical timescales are symmetric under $\theta \to \pi/2 - \theta$ as well.

\end{widetext}

\bibliographystyle{apsrev4-1}

\bibliography{OQN_bib}

\begin{thebibliography}{35}%
\makeatletter
\providecommand \@ifxundefined [1]{%
 \@ifx{#1\undefined}
}%
\providecommand \@ifnum [1]{%
 \ifnum #1\expandafter \@firstoftwo
 \else \expandafter \@secondoftwo
 \fi
}%
\providecommand \@ifx [1]{%
 \ifx #1\expandafter \@firstoftwo
 \else \expandafter \@secondoftwo
 \fi
}%
\providecommand \natexlab [1]{#1}%
\providecommand \enquote  [1]{``#1''}%
\providecommand \bibnamefont  [1]{#1}%
\providecommand \bibfnamefont [1]{#1}%
\providecommand \citenamefont [1]{#1}%
\providecommand \href@noop [0]{\@secondoftwo}%
\providecommand \href [0]{\begingroup \@sanitize@url \@href}%
\providecommand \@href[1]{\@@startlink{#1}\@@href}%
\providecommand \@@href[1]{\endgroup#1\@@endlink}%
\providecommand \@sanitize@url [0]{\catcode `\\12\catcode `\$12\catcode
  `\&12\catcode `\#12\catcode `\^12\catcode `\_12\catcode `\%12\relax}%
\providecommand \@@startlink[1]{}%
\providecommand \@@endlink[0]{}%
\providecommand \url  [0]{\begingroup\@sanitize@url \@url }%
\providecommand \@url [1]{\endgroup\@href {#1}{\urlprefix }}%
\providecommand \urlprefix  [0]{URL }%
\providecommand \Eprint [0]{\href }%
\providecommand \doibase [0]{http://dx.doi.org/}%
\providecommand \selectlanguage [0]{\@gobble}%
\providecommand \bibinfo  [0]{\@secondoftwo}%
\providecommand \bibfield  [0]{\@secondoftwo}%
\providecommand \translation [1]{[#1]}%
\providecommand \BibitemOpen [0]{}%
\providecommand \bibitemStop [0]{}%
\providecommand \bibitemNoStop [0]{.\EOS\space}%
\providecommand \EOS [0]{\spacefactor3000\relax}%
\providecommand \BibitemShut  [1]{\csname bibitem#1\endcsname}%
\let\auto@bib@innerbib\@empty
\bibitem [{\citenamefont {Kadowaki}\ and\ \citenamefont
  {Nishimori}(1998)}]{Nishimori:PRE:1999}%
  \BibitemOpen
  \bibfield  {author} {\bibinfo {author} {\bibfnamefont {T.}~\bibnamefont
  {Kadowaki}}\ and\ \bibinfo {author} {\bibfnamefont {H.}~\bibnamefont
  {Nishimori}},\ }\href {\doibase 10.1103/PhysRevE.58.5355} {\bibfield
  {journal} {\bibinfo  {journal} {Phys. Rev. E}\ }\textbf {\bibinfo {volume}
  {58}},\ \bibinfo {pages} {5355} (\bibinfo {year} {1998})}\BibitemShut
  {NoStop}%
\bibitem [{\citenamefont {Morita}\ and\ \citenamefont
  {Nishimori}(2008)}]{Nishimori:JMP:2008}%
  \BibitemOpen
  \bibfield  {author} {\bibinfo {author} {\bibfnamefont {S.}~\bibnamefont
  {Morita}}\ and\ \bibinfo {author} {\bibfnamefont {H.}~\bibnamefont
  {Nishimori}},\ }\href {\doibase 10.1063/1.2995837} {\bibfield  {journal}
  {\bibinfo  {journal} {Journal of Mathematical Physics}\ }\textbf {\bibinfo
  {volume} {49}},\ \bibinfo {pages} {125210} (\bibinfo {year}
  {2008})}\BibitemShut {NoStop}%
\bibitem [{\citenamefont {Aharonov}\ \emph {et~al.}(2008)\citenamefont
  {Aharonov}, \citenamefont {Van~Dam}, \citenamefont {Kempe}, \citenamefont
  {Landau}, \citenamefont {Lloyd},\ and\ \citenamefont
  {Regev}}]{Lloyd:SIAM:2008}%
  \BibitemOpen
  \bibfield  {author} {\bibinfo {author} {\bibfnamefont {D.}~\bibnamefont
  {Aharonov}}, \bibinfo {author} {\bibfnamefont {W.}~\bibnamefont {Van~Dam}},
  \bibinfo {author} {\bibfnamefont {J.}~\bibnamefont {Kempe}}, \bibinfo
  {author} {\bibfnamefont {Z.}~\bibnamefont {Landau}}, \bibinfo {author}
  {\bibfnamefont {S.}~\bibnamefont {Lloyd}}, \ and\ \bibinfo {author}
  {\bibfnamefont {O.}~\bibnamefont {Regev}},\ }\href@noop {} {\bibfield
  {journal} {\bibinfo  {journal} {SIAM Review}\ }\textbf {\bibinfo {volume}
  {50}},\ \bibinfo {pages} {755} (\bibinfo {year} {2008})}\BibitemShut
  {NoStop}%
\bibitem [{\citenamefont {Boixo}\ \emph {et~al.}(2014)\citenamefont {Boixo},
  \citenamefont {R{\o}nnow}, \citenamefont {Isakov}, \citenamefont {Wang},
  \citenamefont {Wecker}, \citenamefont {Lidar}, \citenamefont {Martinis},\
  and\ \citenamefont {Troyer}}]{Troyer:NatPhys:2014}%
  \BibitemOpen
  \bibfield  {author} {\bibinfo {author} {\bibfnamefont {S.}~\bibnamefont
  {Boixo}}, \bibinfo {author} {\bibfnamefont {T.~F.}\ \bibnamefont
  {R{\o}nnow}}, \bibinfo {author} {\bibfnamefont {S.~V.}\ \bibnamefont
  {Isakov}}, \bibinfo {author} {\bibfnamefont {Z.}~\bibnamefont {Wang}},
  \bibinfo {author} {\bibfnamefont {D.}~\bibnamefont {Wecker}}, \bibinfo
  {author} {\bibfnamefont {D.~A.}\ \bibnamefont {Lidar}}, \bibinfo {author}
  {\bibfnamefont {J.~M.}\ \bibnamefont {Martinis}}, \ and\ \bibinfo {author}
  {\bibfnamefont {M.}~\bibnamefont {Troyer}},\ }\href
  {http://dx.doi.org/10.1038/nphys2900} {\bibfield  {journal} {\bibinfo
  {journal} {Nature Physics}\ }\textbf {\bibinfo {volume} {10}},\ \bibinfo
  {pages} {218} (\bibinfo {year} {2014})}\BibitemShut {NoStop}%
\bibitem [{\citenamefont {Baldassi}\ and\ \citenamefont
  {Zecchina}(2018)}]{Baldassi:PNAS:2018}%
  \BibitemOpen
  \bibfield  {author} {\bibinfo {author} {\bibfnamefont {C.}~\bibnamefont
  {Baldassi}}\ and\ \bibinfo {author} {\bibfnamefont {R.}~\bibnamefont
  {Zecchina}},\ }\href {\doibase 10.1073/pnas.1711456115} {\bibfield  {journal}
  {\bibinfo  {journal} {Proceedings of the National Academy of Sciences}\
  }\textbf {\bibinfo {volume} {115}},\ \bibinfo {pages} {1457} (\bibinfo {year}
  {2018})}\BibitemShut {NoStop}%
\bibitem [{\citenamefont {Biamonte}\ \emph {et~al.}(2017)\citenamefont
  {Biamonte}, \citenamefont {Wittek}, \citenamefont {Pancotti}, \citenamefont
  {Rebentrost}, \citenamefont {Wiebe},\ and\ \citenamefont
  {Lloyd}}]{Biamonte:Nat:2017}%
  \BibitemOpen
  \bibfield  {author} {\bibinfo {author} {\bibfnamefont {J.}~\bibnamefont
  {Biamonte}}, \bibinfo {author} {\bibfnamefont {P.}~\bibnamefont {Wittek}},
  \bibinfo {author} {\bibfnamefont {N.}~\bibnamefont {Pancotti}}, \bibinfo
  {author} {\bibfnamefont {P.}~\bibnamefont {Rebentrost}}, \bibinfo {author}
  {\bibfnamefont {N.}~\bibnamefont {Wiebe}}, \ and\ \bibinfo {author}
  {\bibfnamefont {S.}~\bibnamefont {Lloyd}},\ }\href@noop {} {\bibfield
  {journal} {\bibinfo  {journal} {Nature}\ }\textbf {\bibinfo {volume} {549}},\
  \bibinfo {pages} {195} (\bibinfo {year} {2017})}\BibitemShut {NoStop}%
\bibitem [{\citenamefont {Shor}(1999)}]{Shor:SIAM:1999}%
  \BibitemOpen
  \bibfield  {author} {\bibinfo {author} {\bibfnamefont {P.~W.}\ \bibnamefont
  {Shor}},\ }\href {\doibase 10.1137/S0036144598347011} {\bibfield  {journal}
  {\bibinfo  {journal} {SIAM Review}\ }\textbf {\bibinfo {volume} {41}},\
  \bibinfo {pages} {303} (\bibinfo {year} {1999})}\BibitemShut {NoStop}%
\bibitem [{\citenamefont {Grover}(1997)}]{Grover:PRL:1997}%
  \BibitemOpen
  \bibfield  {author} {\bibinfo {author} {\bibfnamefont {L.~K.}\ \bibnamefont
  {Grover}},\ }\href {\doibase 10.1103/PhysRevLett.79.325} {\bibfield
  {journal} {\bibinfo  {journal} {Phys. Rev. Lett.}\ }\textbf {\bibinfo
  {volume} {79}},\ \bibinfo {pages} {325} (\bibinfo {year} {1997})}\BibitemShut
  {NoStop}%
\bibitem [{\citenamefont {Das}\ and\ \citenamefont
  {Chakrabarti}(2008)}]{Das2008}%
  \BibitemOpen
  \bibfield  {author} {\bibinfo {author} {\bibfnamefont {A.}~\bibnamefont
  {Das}}\ and\ \bibinfo {author} {\bibfnamefont {B.~K.}\ \bibnamefont
  {Chakrabarti}},\ }\href {\doibase 10.1103/RevModPhys.80.1061} {\bibfield
  {journal} {\bibinfo  {journal} {Rev. Mod. Phys.}\ }\textbf {\bibinfo {volume}
  {80}},\ \bibinfo {pages} {1061} (\bibinfo {year} {2008})}\BibitemShut
  {NoStop}%
\bibitem [{\citenamefont {Farhi}\ \emph {et~al.}(2001)\citenamefont {Farhi},
  \citenamefont {Goldstone}, \citenamefont {Gutmann}, \citenamefont {Lapan},
  \citenamefont {Lundgren},\ and\ \citenamefont {Preda}}]{Farhi472}%
  \BibitemOpen
  \bibfield  {author} {\bibinfo {author} {\bibfnamefont {E.}~\bibnamefont
  {Farhi}}, \bibinfo {author} {\bibfnamefont {J.}~\bibnamefont {Goldstone}},
  \bibinfo {author} {\bibfnamefont {S.}~\bibnamefont {Gutmann}}, \bibinfo
  {author} {\bibfnamefont {J.}~\bibnamefont {Lapan}}, \bibinfo {author}
  {\bibfnamefont {A.}~\bibnamefont {Lundgren}}, \ and\ \bibinfo {author}
  {\bibfnamefont {D.}~\bibnamefont {Preda}},\ }\href {\doibase
  10.1126/science.1057726} {\bibfield  {journal} {\bibinfo  {journal}
  {Science}\ }\textbf {\bibinfo {volume} {292}},\ \bibinfo {pages} {472}
  (\bibinfo {year} {2001})}\BibitemShut {NoStop}%
\bibitem [{\citenamefont {Santoro}\ \emph {et~al.}(2002)\citenamefont
  {Santoro}, \citenamefont {Marto{\v{n}}{\'a}k}, \citenamefont {Tosatti},\ and\
  \citenamefont {Car}}]{santoro2002theory}%
  \BibitemOpen
  \bibfield  {author} {\bibinfo {author} {\bibfnamefont {G.~E.}\ \bibnamefont
  {Santoro}}, \bibinfo {author} {\bibfnamefont {R.}~\bibnamefont
  {Marto{\v{n}}{\'a}k}}, \bibinfo {author} {\bibfnamefont {E.}~\bibnamefont
  {Tosatti}}, \ and\ \bibinfo {author} {\bibfnamefont {R.}~\bibnamefont
  {Car}},\ }\href@noop {} {\bibfield  {journal} {\bibinfo  {journal} {Science}\
  }\textbf {\bibinfo {volume} {295}},\ \bibinfo {pages} {2427} (\bibinfo {year}
  {2002})}\BibitemShut {NoStop}%
\bibitem [{\citenamefont {Nishimori}\ and\ \citenamefont
  {Takada}(2017)}]{nishimori2017exponential}%
  \BibitemOpen
  \bibfield  {author} {\bibinfo {author} {\bibfnamefont {H.}~\bibnamefont
  {Nishimori}}\ and\ \bibinfo {author} {\bibfnamefont {K.}~\bibnamefont
  {Takada}},\ }\href@noop {} {\bibfield  {journal} {\bibinfo  {journal}
  {Frontiers in ICT}\ }\textbf {\bibinfo {volume} {4}},\ \bibinfo {pages} {2}
  (\bibinfo {year} {2017})}\BibitemShut {NoStop}%
\bibitem [{\citenamefont {Denchev}\ \emph {et~al.}(2016)\citenamefont
  {Denchev}, \citenamefont {Boixo}, \citenamefont {Isakov}, \citenamefont
  {Ding}, \citenamefont {Babbush}, \citenamefont {Smelyanskiy}, \citenamefont
  {Martinis},\ and\ \citenamefont {Neven}}]{Denchev:PRX:2016}%
  \BibitemOpen
  \bibfield  {author} {\bibinfo {author} {\bibfnamefont {V.~S.}\ \bibnamefont
  {Denchev}}, \bibinfo {author} {\bibfnamefont {S.}~\bibnamefont {Boixo}},
  \bibinfo {author} {\bibfnamefont {S.~V.}\ \bibnamefont {Isakov}}, \bibinfo
  {author} {\bibfnamefont {N.}~\bibnamefont {Ding}}, \bibinfo {author}
  {\bibfnamefont {R.}~\bibnamefont {Babbush}}, \bibinfo {author} {\bibfnamefont
  {V.}~\bibnamefont {Smelyanskiy}}, \bibinfo {author} {\bibfnamefont
  {J.}~\bibnamefont {Martinis}}, \ and\ \bibinfo {author} {\bibfnamefont
  {H.}~\bibnamefont {Neven}},\ }\href {\doibase 10.1103/PhysRevX.6.031015}
  {\bibfield  {journal} {\bibinfo  {journal} {Phys. Rev. X}\ }\textbf {\bibinfo
  {volume} {6}},\ \bibinfo {pages} {031015} (\bibinfo {year}
  {2016})}\BibitemShut {NoStop}%
\bibitem [{\citenamefont {Breuer}\ and\ \citenamefont
  {Petruccione}(2002)}]{BreuerP:2002}%
  \BibitemOpen
  \bibfield  {author} {\bibinfo {author} {\bibfnamefont {H.~P.}\ \bibnamefont
  {Breuer}}\ and\ \bibinfo {author} {\bibfnamefont {F.}~\bibnamefont
  {Petruccione}},\ }\href@noop {} {\emph {\bibinfo {title} {The theory of open
  quantum systems}}}\ (\bibinfo  {publisher} {Oxford University Press},\
  \bibinfo {address} {Great Clarendon Street},\ \bibinfo {year}
  {2002})\BibitemShut {NoStop}%
\bibitem [{\citenamefont {Diehl}\ \emph {et~al.}(2008)\citenamefont {Diehl},
  \citenamefont {Micheli}, \citenamefont {Kantian}, \citenamefont {Kraus},
  \citenamefont {B{\"u}chler},\ and\ \citenamefont
  {Zoller}}]{diehl2008quantum}%
  \BibitemOpen
  \bibfield  {author} {\bibinfo {author} {\bibfnamefont {S.}~\bibnamefont
  {Diehl}}, \bibinfo {author} {\bibfnamefont {A.}~\bibnamefont {Micheli}},
  \bibinfo {author} {\bibfnamefont {A.}~\bibnamefont {Kantian}}, \bibinfo
  {author} {\bibfnamefont {B.}~\bibnamefont {Kraus}}, \bibinfo {author}
  {\bibfnamefont {H.}~\bibnamefont {B{\"u}chler}}, \ and\ \bibinfo {author}
  {\bibfnamefont {P.}~\bibnamefont {Zoller}},\ }\href
  {http://dx.doi.org/10.1038/nphys1073} {\bibfield  {journal} {\bibinfo
  {journal} {Nature Physics}\ }\textbf {\bibinfo {volume} {4}},\ \bibinfo
  {pages} {878} (\bibinfo {year} {2008})}\BibitemShut {NoStop}%
\bibitem [{\citenamefont {Müller}\ \emph {et~al.}(2012)\citenamefont
  {Müller}, \citenamefont {Diehl}, \citenamefont {Pupillo},\ and\
  \citenamefont {Zoller}}]{MULLER20121}%
  \BibitemOpen
  \bibfield  {author} {\bibinfo {author} {\bibfnamefont {M.}~\bibnamefont
  {Müller}}, \bibinfo {author} {\bibfnamefont {S.}~\bibnamefont {Diehl}},
  \bibinfo {author} {\bibfnamefont {G.}~\bibnamefont {Pupillo}}, \ and\
  \bibinfo {author} {\bibfnamefont {P.}~\bibnamefont {Zoller}},\ }in\ \href
  {\doibase https://doi.org/10.1016/B978-0-12-396482-3.00001-6} {\emph
  {\bibinfo {booktitle} {Advances in Atomic, Molecular, and Optical
  Physics}}},\ \bibinfo {series} {Advances In Atomic, Molecular, and Optical
  Physics}, Vol.~\bibinfo {volume} {61},\ \bibinfo {editor} {edited by\
  \bibinfo {editor} {\bibfnamefont {P.}~\bibnamefont {Berman}}, \bibinfo
  {editor} {\bibfnamefont {E.}~\bibnamefont {Arimondo}}, \ and\ \bibinfo
  {editor} {\bibfnamefont {C.}~\bibnamefont {Lin}}}\ (\bibinfo  {publisher}
  {Academic Press},\ \bibinfo {year} {2012})\ pp.\ \bibinfo {pages} {1 --
  80}\BibitemShut {NoStop}%
\bibitem [{\citenamefont {Verstraete}\ \emph {et~al.}(2009)\citenamefont
  {Verstraete}, \citenamefont {Wolf},\ and\ \citenamefont
  {Cirac}}]{Verstraete:NatPhys:2009}%
  \BibitemOpen
  \bibfield  {author} {\bibinfo {author} {\bibfnamefont {F.}~\bibnamefont
  {Verstraete}}, \bibinfo {author} {\bibfnamefont {M.~M.}\ \bibnamefont
  {Wolf}}, \ and\ \bibinfo {author} {\bibfnamefont {J.~I.}\ \bibnamefont
  {Cirac}},\ }\href
  {http://www.nature.com/nphys/journal/v5/n9/abs/nphys1342.html} {\bibfield
  {journal} {\bibinfo  {journal} {Nature Physics}\ }\textbf {\bibinfo {volume}
  {5}},\ \bibinfo {pages} {633} (\bibinfo {year} {2009})}\BibitemShut {NoStop}%
\bibitem [{\citenamefont {Alicki}(1976)}]{alicki1976}%
  \BibitemOpen
  \bibfield  {author} {\bibinfo {author} {\bibfnamefont {R.}~\bibnamefont
  {Alicki}},\ }\href@noop {} {\bibfield  {journal} {\bibinfo  {journal}
  {Reports on Mathematical Physics}\ }\textbf {\bibinfo {volume} {10}},\
  \bibinfo {pages} {249} (\bibinfo {year} {1976})}\BibitemShut {NoStop}%
\bibitem [{\citenamefont {Olmos}\ \emph {et~al.}(2014)\citenamefont {Olmos},
  \citenamefont {Lesanovsky},\ and\ \citenamefont {Garrahan}}]{Olmos2014}%
  \BibitemOpen
  \bibfield  {author} {\bibinfo {author} {\bibfnamefont {B.}~\bibnamefont
  {Olmos}}, \bibinfo {author} {\bibfnamefont {I.}~\bibnamefont {Lesanovsky}}, \
  and\ \bibinfo {author} {\bibfnamefont {J.~P.}\ \bibnamefont {Garrahan}},\
  }\href {\doibase 10.1103/PhysRevE.90.042147} {\bibfield  {journal} {\bibinfo
  {journal} {Phys. Rev. E}\ }\textbf {\bibinfo {volume} {90}},\ \bibinfo
  {pages} {042147} (\bibinfo {year} {2014})}\BibitemShut {NoStop}%
\bibitem [{\citenamefont {Marcuzzi}\ \emph {et~al.}(2014)\citenamefont
  {Marcuzzi}, \citenamefont {Schick}, \citenamefont {Olmos},\ and\
  \citenamefont {Lesanovsky}}]{Marcuzzi:JPA:2014}%
  \BibitemOpen
  \bibfield  {author} {\bibinfo {author} {\bibfnamefont {M.}~\bibnamefont
  {Marcuzzi}}, \bibinfo {author} {\bibfnamefont {J.}~\bibnamefont {Schick}},
  \bibinfo {author} {\bibfnamefont {B.}~\bibnamefont {Olmos}}, \ and\ \bibinfo
  {author} {\bibfnamefont {I.}~\bibnamefont {Lesanovsky}},\ }\href
  {http://stacks.iop.org/1751-8121/47/i=48/a=482001} {\bibfield  {journal}
  {\bibinfo  {journal} {Journal of Physics A: Mathematical and Theoretical}\
  }\textbf {\bibinfo {volume} {47}},\ \bibinfo {pages} {482001} (\bibinfo
  {year} {2014})}\BibitemShut {NoStop}%
\bibitem [{\citenamefont {Hopfield}(1982)}]{Hopfield:1982}%
  \BibitemOpen
  \bibfield  {author} {\bibinfo {author} {\bibfnamefont {J.~J.}\ \bibnamefont
  {Hopfield}},\ }\href@noop {} {\bibfield  {journal} {\bibinfo  {journal}
  {PNAS}\ }\textbf {\bibinfo {volume} {79}},\ \bibinfo {pages} {2554} (\bibinfo
  {year} {1982})}\BibitemShut {NoStop}%
\bibitem [{\citenamefont {Rotondo}\ \emph {et~al.}(2018)\citenamefont
  {Rotondo}, \citenamefont {Marcuzzi}, \citenamefont {Garrahan}, \citenamefont
  {Lesanovsky},\ and\ \citenamefont {M\"uller}}]{Rotondo:JPA:2018}%
  \BibitemOpen
  \bibfield  {author} {\bibinfo {author} {\bibfnamefont {P.}~\bibnamefont
  {Rotondo}}, \bibinfo {author} {\bibfnamefont {M.}~\bibnamefont {Marcuzzi}},
  \bibinfo {author} {\bibfnamefont {J.~P.}\ \bibnamefont {Garrahan}}, \bibinfo
  {author} {\bibfnamefont {I.}~\bibnamefont {Lesanovsky}}, \ and\ \bibinfo
  {author} {\bibfnamefont {M.}~\bibnamefont {M\"uller}},\ }\href
  {http://stacks.iop.org/1751-8121/51/i=11/a=115301} {\bibfield  {journal}
  {\bibinfo  {journal} {Journal of Physics A: Mathematical and Theoretical}\
  }\textbf {\bibinfo {volume} {51}},\ \bibinfo {pages} {115301} (\bibinfo
  {year} {2018})}\BibitemShut {NoStop}%
\bibitem [{\citenamefont {Mezard}\ and\ \citenamefont
  {Montanari}(2009)}]{Montanari:book}%
  \BibitemOpen
  \bibfield  {author} {\bibinfo {author} {\bibfnamefont {M.}~\bibnamefont
  {Mezard}}\ and\ \bibinfo {author} {\bibfnamefont {A.}~\bibnamefont
  {Montanari}},\ }\href@noop {} {\emph {\bibinfo {title} {Information, physics,
  and computation}}}\ (\bibinfo  {publisher} {Oxford University Press},\
  \bibinfo {year} {2009})\BibitemShut {NoStop}%
\bibitem [{\citenamefont {Lindblad}(1976)}]{Lindblad76}%
  \BibitemOpen
  \bibfield  {author} {\bibinfo {author} {\bibfnamefont {G.}~\bibnamefont
  {Lindblad}},\ }\href {\doibase 10.1007/BF01608499} {\bibfield  {journal}
  {\bibinfo  {journal} {Communications in Mathematical Physics}\ }\textbf
  {\bibinfo {volume} {48}},\ \bibinfo {pages} {119} (\bibinfo {year}
  {1976})}\BibitemShut {NoStop}%
\bibitem [{\citenamefont {Breuer}\ and\ \citenamefont
  {Petruccione}(2007)}]{Breuer_P}%
  \BibitemOpen
  \bibfield  {author} {\bibinfo {author} {\bibfnamefont {H.}~\bibnamefont
  {Breuer}}\ and\ \bibinfo {author} {\bibfnamefont {F.}~\bibnamefont
  {Petruccione}},\ }\href {https://books.google.co.uk/books?id=DkcJPwAACAAJ}
  {\emph {\bibinfo {title} {The Theory of Open Quantum Systems}}}\ (\bibinfo
  {publisher} {OUP Oxford},\ \bibinfo {year} {2007})\BibitemShut {NoStop}%
\bibitem [{\citenamefont {Castelnovo}\ \emph {et~al.}(2005)\citenamefont
  {Castelnovo}, \citenamefont {Chamon}, \citenamefont {Mudry},\ and\
  \citenamefont {Pujol}}]{Castelnovo:AnnPhys:2005}%
  \BibitemOpen
  \bibfield  {author} {\bibinfo {author} {\bibfnamefont {C.}~\bibnamefont
  {Castelnovo}}, \bibinfo {author} {\bibfnamefont {C.}~\bibnamefont {Chamon}},
  \bibinfo {author} {\bibfnamefont {C.}~\bibnamefont {Mudry}}, \ and\ \bibinfo
  {author} {\bibfnamefont {P.}~\bibnamefont {Pujol}},\ }\href {\doibase
  https://doi.org/10.1016/j.aop.2005.01.006} {\bibfield  {journal} {\bibinfo
  {journal} {Annals of Physics}\ }\textbf {\bibinfo {volume} {318}},\ \bibinfo
  {pages} {316 } (\bibinfo {year} {2005})}\BibitemShut {NoStop}%
\bibitem [{SM()}]{SM}%
  \BibitemOpen
  \href@noop {} {\bibinfo  {journal} {see supplemental material}\ }\BibitemShut
  {NoStop}%
\bibitem [{\citenamefont {Amit}\ \emph
  {et~al.}(1985{\natexlab{a}})\citenamefont {Amit}, \citenamefont {Gutfreund},\
  and\ \citenamefont {Sompolinsky}}]{AmitGS:1985a}%
  \BibitemOpen
\bibfield  {journal} {  }\bibfield  {author} {\bibinfo {author} {\bibfnamefont
  {D.~J.}\ \bibnamefont {Amit}}, \bibinfo {author} {\bibfnamefont
  {H.}~\bibnamefont {Gutfreund}}, \ and\ \bibinfo {author} {\bibfnamefont
  {H.}~\bibnamefont {Sompolinsky}},\ }\href@noop {} {\bibfield  {journal}
  {\bibinfo  {journal} {Phys. Rev. A}\ }\textbf {\bibinfo {volume} {32}},\
  \bibinfo {pages} {1007} (\bibinfo {year} {1985}{\natexlab{a}})}\BibitemShut
  {NoStop}%
\bibitem [{\citenamefont {Amit}\ \emph
  {et~al.}(1985{\natexlab{b}})\citenamefont {Amit}, \citenamefont {Gutfreund},\
  and\ \citenamefont {Sompolinsky}}]{AmitGS:1985b}%
  \BibitemOpen
  \bibfield  {author} {\bibinfo {author} {\bibfnamefont {D.~J.}\ \bibnamefont
  {Amit}}, \bibinfo {author} {\bibfnamefont {H.}~\bibnamefont {Gutfreund}}, \
  and\ \bibinfo {author} {\bibfnamefont {H.}~\bibnamefont {Sompolinsky}},\
  }\href {\doibase 10.1103/PhysRevLett.55.1530} {\bibfield  {journal} {\bibinfo
   {journal} {Phys. Rev. Lett.}\ }\textbf {\bibinfo {volume} {55}},\ \bibinfo
  {pages} {1530} (\bibinfo {year} {1985}{\natexlab{b}})}\BibitemShut {NoStop}%
\bibitem [{\citenamefont {M{\'e}zard}\ \emph {et~al.}(1990)\citenamefont
  {M{\'e}zard}, \citenamefont {Parisi},\ and\ \citenamefont
  {Virasoro}}]{Mezard:book}%
  \BibitemOpen
  \bibfield  {author} {\bibinfo {author} {\bibfnamefont {M.}~\bibnamefont
  {M{\'e}zard}}, \bibinfo {author} {\bibfnamefont {G.}~\bibnamefont {Parisi}},
  \ and\ \bibinfo {author} {\bibfnamefont {M.-A.}\ \bibnamefont {Virasoro}},\
  }\href@noop {} {\emph {\bibinfo {title} {Spin glass theory and beyond.}}}\
  (\bibinfo  {publisher} {World Scientific Publishing Co., Inc., Pergamon
  Press},\ \bibinfo {year} {1990})\BibitemShut {NoStop}%
\bibitem [{\citenamefont {Gopalakrishnan}\ \emph {et~al.}(2011)\citenamefont
  {Gopalakrishnan}, \citenamefont {Lev},\ and\ \citenamefont
  {Goldbart}}]{Goldbart:PRL:2011}%
  \BibitemOpen
  \bibfield  {author} {\bibinfo {author} {\bibfnamefont {S.}~\bibnamefont
  {Gopalakrishnan}}, \bibinfo {author} {\bibfnamefont {B.}~\bibnamefont {Lev}},
  \ and\ \bibinfo {author} {\bibfnamefont {P.}~\bibnamefont {Goldbart}},\
  }\href@noop {} {\bibfield  {journal} {\bibinfo  {journal} {Phys. Rev. Lett.}\
  }\textbf {\bibinfo {volume} {107}},\ \bibinfo {pages} {277201} (\bibinfo
  {year} {2011})}\BibitemShut {NoStop}%
\bibitem [{\citenamefont {Strack}\ and\ \citenamefont
  {Sachdev}(2011)}]{Strack:PRL:2011}%
  \BibitemOpen
  \bibfield  {author} {\bibinfo {author} {\bibfnamefont {P.}~\bibnamefont
  {Strack}}\ and\ \bibinfo {author} {\bibfnamefont {S.}~\bibnamefont
  {Sachdev}},\ }\href@noop {} {\bibfield  {journal} {\bibinfo  {journal} {Phys.
  Rev. Lett.}\ }\textbf {\bibinfo {volume} {107}},\ \bibinfo {pages} {277202}
  (\bibinfo {year} {2011})}\BibitemShut {NoStop}%
\bibitem [{\citenamefont {Rotondo}\ \emph
  {et~al.}(2015{\natexlab{a}})\citenamefont {Rotondo}, \citenamefont
  {Cosentino~Lagomarsino},\ and\ \citenamefont {Viola}}]{Rotondo:PRL:2015}%
  \BibitemOpen
  \bibfield  {author} {\bibinfo {author} {\bibfnamefont {P.}~\bibnamefont
  {Rotondo}}, \bibinfo {author} {\bibfnamefont {M.}~\bibnamefont
  {Cosentino~Lagomarsino}}, \ and\ \bibinfo {author} {\bibfnamefont
  {G.}~\bibnamefont {Viola}},\ }\href {\doibase 10.1103/PhysRevLett.114.143601}
  {\bibfield  {journal} {\bibinfo  {journal} {Phys. Rev. Lett.}\ }\textbf
  {\bibinfo {volume} {114}},\ \bibinfo {pages} {143601} (\bibinfo {year}
  {2015}{\natexlab{a}})}\BibitemShut {NoStop}%
\bibitem [{\citenamefont {Rotondo}\ \emph
  {et~al.}(2015{\natexlab{b}})\citenamefont {Rotondo}, \citenamefont {Tesio},\
  and\ \citenamefont {Caracciolo}}]{Rotondo:PRB:2015}%
  \BibitemOpen
  \bibfield  {author} {\bibinfo {author} {\bibfnamefont {P.}~\bibnamefont
  {Rotondo}}, \bibinfo {author} {\bibfnamefont {E.}~\bibnamefont {Tesio}}, \
  and\ \bibinfo {author} {\bibfnamefont {S.}~\bibnamefont {Caracciolo}},\
  }\href {\doibase 10.1103/PhysRevB.91.014415} {\bibfield  {journal} {\bibinfo
  {journal} {Phys. Rev. B}\ }\textbf {\bibinfo {volume} {91}},\ \bibinfo
  {pages} {014415} (\bibinfo {year} {2015}{\natexlab{b}})}\BibitemShut
  {NoStop}%
\bibitem [{\citenamefont {Vaidya}\ \emph {et~al.}(2018)\citenamefont {Vaidya},
  \citenamefont {Guo}, \citenamefont {Kroeze}, \citenamefont {Ballantine},
  \citenamefont {Koll\'ar}, \citenamefont {Keeling},\ and\ \citenamefont
  {Lev}}]{Lev:PRX:2018}%
  \BibitemOpen
  \bibfield  {author} {\bibinfo {author} {\bibfnamefont {V.~D.}\ \bibnamefont
  {Vaidya}}, \bibinfo {author} {\bibfnamefont {Y.}~\bibnamefont {Guo}},
  \bibinfo {author} {\bibfnamefont {R.~M.}\ \bibnamefont {Kroeze}}, \bibinfo
  {author} {\bibfnamefont {K.~E.}\ \bibnamefont {Ballantine}}, \bibinfo
  {author} {\bibfnamefont {A.~J.}\ \bibnamefont {Koll\'ar}}, \bibinfo {author}
  {\bibfnamefont {J.}~\bibnamefont {Keeling}}, \ and\ \bibinfo {author}
  {\bibfnamefont {B.~L.}\ \bibnamefont {Lev}},\ }\href {\doibase
  10.1103/PhysRevX.8.011002} {\bibfield  {journal} {\bibinfo  {journal} {Phys.
  Rev. X}\ }\textbf {\bibinfo {volume} {8}},\ \bibinfo {pages} {011002}
  (\bibinfo {year} {2018})}\BibitemShut {NoStop}%
\end{thebibliography}%

\end{document}